\documentclass[12pt]{article}

\usepackage{enumitem}
\usepackage{graphicx}
\usepackage{float}
\usepackage{cite}
\usepackage{amsfonts}
\usepackage{amssymb}
\usepackage{amsmath}
\usepackage{xcolor}

\def\gtwid{\mathrel{\raise.3ex\hbox{$>$\kern-.75em\lower1ex\hbox{$\sim$}}}}
\def\ltwid{\mathrel{\raise.3ex\hbox{$<$\kern-.75em\lower1ex\hbox{$\sim$}}}}
\def\square{\kern1pt\vbox{\hrule height 1.2pt\hbox{\vrule width 1.2pt\hskip 3pt
  \vbox{\vskip 6pt}\hskip 3pt\vrule width 0.6pt}\hrule height 0.6pt}\kern1pt}

\begin{document}

\begin{titlepage}

\begin{flushright}
CCTP-2026-17 \\
ITCP/2026/17 \\
UFIFT-QG-26-08
\end{flushright}

\vskip 1cm

\begin{center}
{\bf Consistency Checks for $\Lambda \neq 0$ Quantum Cosmology}
\end{center}

\vskip 1cm

\begin{center}
S. P. Miao$^{1\star}$, N. C. Tsamis$^{2\dagger}$ and 
R. P. Woodard$^{3\ddagger}$
\end{center}

\vskip 0.5cm

\begin{center}
\it{$^{1}$ Department of Physics, National Cheng Kung University, \\
No. 1 University Road, Tainan City 70101, TAIWAN}
\end{center}

\begin{center}
\it{$^{2}$ Institute of Theoretical Physics \& Computational Physics, \\
Department of Physics, University of Crete, \\
GR-700 13 Heraklion, HELLAS}
\end{center}

\begin{center}
\it{$^{3}$ Department of Physics, University of Florida,\\
Gainesville, FL 32611, UNITED STATES}
\end{center}

\vspace{0cm}

\begin{center}
ABSTRACT
\end{center}
In this paper we improve and further analyze the recently 
derived quantum gravity equations which apply in accelerating 
cosmological spacetimes and whose solutions re-sum leading 
logarithms from all orders of perturbation theory. We show 
the consistency of these upgraded equations against 1-loop 
non-trivial exact perturbative computations. The upgraded
equations retain their remarkably simple form. 

\begin{flushleft}
\end{flushleft}

\vskip 0.5cm

\begin{flushleft}
$^{\star}$ e-mail: spmiao5@mail.ncku.edu.tw \\
$^{\dagger}$ e-mail: tsamis@physics.uoc.gr \\
$^{\ddagger}$ e-mail: woodard@phys.ufl.edu
\end{flushleft}

\end{titlepage}

\section{Prologue}

A standard example of an accelerating spacetime which
seems to be of physical significance during early cosmological
evolution is the primordial inflationary epoch defined by
$H > 0$ with $0 \leq \epsilon < 1$ \cite{Geshnizjani:2011dk}.
Any cosmological geometry is characterized by a scale factor
$a(t)$ and its two first time derivatives, the Hubble parameter
$H(t)$ and the 1st slow roll parameter $\epsilon(t)$:
\footnote{It is more often than not convenient to employ
conformal instead of co-moving coordinates:
$ds^2 \!=\!
-dt^2 + a^2(t) \, d{\mathbf{x}} \cdot d{\mathbf{x}}
=
a^2(\eta) \big[\! -d\eta^2 + a^2(\eta) \,
d{\mathbf{x}} \cdot d{\mathbf{x}} \big]$, with
$t$ the co-moving time and $\eta$ the conformal
time.}
\begin{equation}
ds^2 = - dt^2 + a^2(t) \, d{\mathbf{x}} \cdot d{\mathbf{x}}
\qquad \Longrightarrow \qquad
H(t) \equiv \frac{\dot{a}}{a}
\quad, \quad
\epsilon(t) \equiv - \frac{\dot{H}}{H^2}
\; . \label{geometry}
\end{equation}
However, there is no agreement on the cause of  primordial
inflation \cite{Ijjas:2013vea,Guth:2013sya,Linde:2014nna},
nor is it understood how the parameters that characterize
the current universe assumed their sometimes unexpected
values. A reasonable conjecture is that the answer may
involve quantum gravitational corrections, which grew to
become non-perturbatively strong during a prolonged era
of inflation \cite{Tsamis:1996qm, Tsamis:2011ep}; after
all, it is gravitation which plays the dominant role in
shaping cosmological evolution.
\footnote{It goes without saying that the study of quantum
physics in de Sitter spacetime, the maximally symmetric
example of an accelerating spacetime, already has a long
history \cite{Grishchuk:1977zz,
Myhrvold:1983hx,
Ford:1984hs,
Allen:1985wd,
Antoniadis:1986sb,
Allen:1986tt,
Floratos:1987ek,
Higuchi:1991tm,
Tsamis:1992xa,
Dolgov:1994cq,
Tsamis:2005hd,
Polyakov:2007mm,
Giddings:2007nu,
Perez-Nadal:2007yxe,
Burgess:2010dd,
Polyakov:2012uc,
Marolf:2012kh,
Anderson:2013ila,
Frob:2013ht,
Anninos:2014lwa,
Frob:2014zka,
Dvali:2014gua,
Burgess:2015ajz,
Brandenberger:2018fdd,
Baumgart:2019clc,
Brahma:2021mng,
Colas:2022hlq,
Cable:2023gdz,
Burgess:2024eng,
Anninos:2024fty,
Brahma:2024yor,
Sloth:2025nan,
Ansari:2025nng,
Ahmadi:2025oon,
Kaplanek:2025moq,
Prokopec:2025jrd}.}

Although quantum gravity is not perturbatively renormalizable,
it can still be used reliably as a quantum low-energy effective
field theory in the generic sense of Weinberg \cite{DHoker:1994rdl}
as, for instance, done by Donoghue
\cite{Donoghue:1993eb,Donoghue:1994dn,Donoghue:2017ovt}.
Quantum gravity was much stronger during primordial inflation
than today because the relevant loop-counting parameter
$\tfrac{\hbar G H^2(t)}{c^5}$ -- which is only about $10^{-122}$
currently -- may have been as large as $10^{-11}$ towards the end
of inflation and even bigger before. While this is small enough
to make perturbation theory valid, it is large enough to produce
detectable effects \cite{Starobinsky:1979ty,Mukhanov:1981xt}.

These effects survive to the current epoch because light,
minimally coupled scalars and gravitons fossilize when
their physical wave numbers $\tfrac{k}{a(t)}$ fall below
the inverse Hubble length $\tfrac{H(t)}{c}$
\cite{Mukhanov:1990me}. Since this horizon crossing continues
as long as accelerated expansion persists, loop corrections
that would be constant in flat space tend to grow with time,
and sometimes also with space. For instance, the vacuum
polarization in de Sitter spacetime \cite{Leonard:2013xsa}
from a loop of gravitons changes the electric field strength
$F_{0i}(t,\bf{x})$ of plane wave photons \cite{Wang:2014tza}
and the Coulomb $\Phi(t, r)$ potential \cite{Glavan:2013jca}
to:
\footnote{The $G \, r^{-2}\!$ correction in (\ref{Coulomb})
is the cosmological descendant of an effect long known from
flat space background \cite{Radkowski:1970ovx}.}
\begin{eqnarray}
F_{0i}(t,{\bf x}) &\!\!\! = \!\!\!&
F^{\rm tree}_{0i}(t,{\bf x}) \, \Bigl\{
1 + \tfrac{2 \hbar G H^2}{\pi c^5} \ln[a(t)] + \dots \Bigr\}
\; , \label{photons} \\
\Phi(t,r) &\!\!\! = \!\!\!&
\tfrac{Q}{4\pi \varepsilon_0 \, a(t) \, r} \, \Bigl\{
1 + \tfrac{2 \hbar G}{3 \pi c^3 a^2(t) \, r^2}
+ \tfrac{2 \hbar G H^2}{\pi c^5} \ln[ \tfrac{a(t) H r}{c}]
+ \dots \Bigr\}
\; . \label{Coulomb}
\end{eqnarray}
The large logarithmic terms proportional to $G H^2$ in
(\ref{photons}-\ref{Coulomb}) are {\it new} effects that
derive from inflationary gravitons.
\footnote{A number of similar logarithmic terms in de Sitter
background can be found in \cite{Miao:2005am,Miao:2006gj,
Tan:2021ibs,Tan:2021lza,Glavan:2021adm,Tan:2022xpn}.}

By studying non-linear sigma models \cite{Miao:2021gic},
and proceeding to scalar loop corrections to gravity
\cite{Miao:2024nsz}, it can be shown that the large
logarithmic terms can be {\it resummed} by combining
a variant of Starobinsky's original stochastic formalism
\cite{Starobinsky:1986fx} with a variant of the
renormalization group. For instance, scalars on de Sitter
spacetime change the electric components of the Weyl tensor
$C_{0i0j}$ for plane wave gravitons, and the Newtonian
potential $\Psi$ to \cite{Miao:2024nsz,Miao:2024atw}:
\begin{equation}
C_{0i0j} =
C^{\rm tree}_{0i0j} \, \Bigl\{
1 - \tfrac{3 \hbar G H^2}{10 \pi c^5} \ln[a] + \dots\Bigr\}
\; \longrightarrow \;
C^{\rm tree}_{0i0j} \!\times\!
[ a(t)]^{-\frac{3 \hbar G H^2}{10 \pi c^5}}
\, , \qquad\qquad \label{Weyl}
\end{equation}
\vspace{-0.5cm}
\begin{equation}
\Psi =
\tfrac{G M}{a r} \, \Bigl\{
1 + \tfrac{\hbar G}{20 \pi c^3 a^2 r^2}
- \tfrac{3 \hbar G H^2}{10 \pi c^5} \ln[\tfrac{a H r}{c}]
+ \dots \Bigr\}
\; \longrightarrow \;
\tfrac{G M}{a r} \!\times\!
[ \tfrac{a(t) H r}{c}]^{-\frac{3 \hbar G H^2}{10 \pi c^5}}
, \label{Newtonscalar}
\end{equation}
where the results on the far right represent the
non-perturbative resummations.

Most recently, a set of rather simple stochastic field
equations has been suggested for pure quantum gravity 
in a fixed gauge \cite{Miao:2024shs,Miao:2025gzm}. Unlike
previous perturbative loop computations, these equations
can, in principle, be solved {\it non-perturbatively} as
long as the expansion rate changes slowly. Additionally,
this set of equations in its perturbative correspondence 
limit predicts changes to the expansion rate \cite{Miao:2025bmd}. 
On the other hand, upon being subjected to further perturbative 
checks, the same set of equations failed to achieve consistency.
The reasons for this discrepancy, the requisite changes to
restore agreement, and the verification of the latter in
specific 1-loop perturbative limits are the main subject 
of this work.

\newpage

\section{Quantum Gravity}

The general $D$-dimensional gravitational effective theory
satisfying general coordinate invariance is an infinite
series of terms with increasing canonical dimensionality.
For cosmology the two lowest terms suffice:
\footnote{Hellenic indices take on spacetime values
while Latin indices take on space values. Our metric
tensor $g_{\mu\nu}$ has spacelike signature
$( - \, + \, + \, +)$ and our curvature tensor equals
$R^{\alpha}_{~ \beta \mu \nu} \equiv
\Gamma^{\alpha}_{~ \nu \beta , \mu} +
\Gamma^{\alpha}_{~ \mu \rho} \,
\Gamma^{\rho}_{~ \nu \beta} -
(\mu \leftrightarrow \nu)$.}
\begin{equation}
{\mathcal L}_{inv} =
\frac{1}{\kappa^2} \Big[ \!-\! (D \!-\! 2) \Lambda + R \, \Big] \sqrt{-g}
\; , \label{Linv}
\end{equation}
where the two parameters are Newton's constant $G$ and the
cosmological constant $\Lambda$:
\footnote{$\Lambda$ is taken to be ``large'' and positive.
Here ``large'' means a $\Lambda$ corresponding to a scale
$M$ which can be as high as $10^{18} \, GeV$.}
\begin{equation}
\kappa^2 \equiv 16 \pi G
\quad , \quad
\Lambda \equiv (D \!-\! 1) H^2
\; . \label{parameters}
\end{equation}
The full equations emanating from (\ref{Linv}) are:
\begin{equation}
R_{\mu\nu} - \frac12 R \, g_{\mu\nu}
+ \frac12 (D \!-\! 2)(D \!-\! 1) H^2 g_{\mu\nu} = 0
\; . \label{eom}
\end{equation}
In terms of the full metric $g_{\mu\nu}$, the conformally
rescaled metric ${\widetilde g}_{\mu\nu}$ and the graviton
field $h_{\mu\nu}$ are defined thusly:
\begin{equation}
g_{\mu\nu} \equiv
a^2 {\widetilde g}_{\mu\nu} \equiv
a^2 \big[ \eta_{\mu\nu} + \kappa h_{\mu\nu} \big]
\; . \label{metrics}
\end{equation}

In view of (\ref{metrics}) the Lagrangian (\ref{Linv})
can be written as follows \cite{Tsamis:1992xa}:
\begin{eqnarray}
{\mathcal L}_{inv} &\!\!\!=\!\!\!&
a^{D-2} \sqrt{-{\widetilde g}} \, {\widetilde g}^{\alpha\beta}
{\widetilde g}^{\rho\sigma} {\widetilde g}^{\mu\nu} 
\Big\{ 
\tfrac12 h_{\alpha\rho, \mu} h_{\nu\sigma, \beta} 
- \tfrac12 h_{\alpha\beta, \rho} h_{\sigma\mu, \nu}
+ \tfrac14 h_{\alpha\beta, \rho} h_{\mu\nu, \sigma}
\nonumber \\
& \mbox{} & 
- \tfrac14 h_{\alpha\rho, \mu} h_{\beta\sigma, \nu}
\Big\}  
+ (\tfrac{D}{2} - 1) a^{D-1} H 
\sqrt{-{\widetilde g}} \, {\widetilde g}^{\rho\sigma}
{\widetilde g}^{\mu\nu}
h_{\rho\sigma, \mu} h_{\nu 0}
\; , \label{Linv2}
\end{eqnarray}
which is the form of $\mathcal{L}_{inv}$ we shall use
thereafter.

\subsection{Degrees of Freedom}

The graviton field variables $A_{\alpha\beta}, B_{\alpha}, C$ are
defined as follows:
\begin{equation}
\kappa \, h_{\alpha\beta}
\equiv 
A_{\alpha\beta}
+ u_{(\alpha} B_{\beta)}
+ ( u_{\alpha} u_{\beta} \!+\! \overline{\gamma}_{\alpha\beta} ) C
\quad , \quad
u^{\alpha} A_{\alpha\beta} = u^{\alpha} B_{\alpha} = 0
\; , \label{ABC}
\end{equation}
where $A_{\alpha\beta}$ contains the dynamical degrees 
of freedom, $B_{\alpha}$ is canonically associated with 
the momentum constraints of general relativity and physically 
represents the strictly general relativistic potentials, 
and $C$ is canonically associated with the Hamiltonian 
constraint of general relativity and physically represents 
the Newtonian potential; in addition, $C$ is the constrained 
field variable in which any dynamical changes of the 
geometrical background are imprinted.

The geometrical variables are those of the Arnowitt-Deser-Misner
(ADM) spatial-temporal decomposition of the conformally re-scaled 
metric \cite{Arnowitt:1962hi}:
\begin{eqnarray}
d{\widetilde s}^{\,\,\! 2} 
& \!\!\!=\!\!\! & 
- N^2 d\eta^2 + \gamma_{ij} (dx^i - N^i d\eta) (dx^j - N^j d\eta)
\; , \label{3+1a} \\
& \!\!\!=\!\!\! & 
(- N^2 \!+\! \gamma_{ij} N^i N^j) \, d\eta^2
- 2 \gamma_{ij} N^j d\eta \, dx^i + \gamma_{ij} dx^i dx^j
\; , \label{3+1b}
\end{eqnarray}
where $N$ is the lapse function, $N^i$ is the shift function, 
and $\gamma_{ij}$ is the spatial metric.
\footnote{The four-vector $u_{\mu}$ -- the ``temporal part'' --
is defined in the Appendix, where the ADM decomposition basics 
can be viewed.}

The field variables $A_{ij}, B_i, C$ are related to the 
geometrical variables $\gamma_{ij}, N^i, N$ thusly:
\begin{equation}
N^2 = \frac{1}{1 + C + \tfrac14 B_i B_i}
\;\; , \;\;
N^i = \frac{\tfrac12 B_i}{\sqrt{1 + C + \tfrac14 B_i B_i}}
\;\; , \;\;
\gamma_{ij} = \frac{\delta_{ij} + A_{ij}}{1 - C}
\; . \label{relation} 
\end{equation}
The remaining variables can be solved for: 
\begin{eqnarray}
u^{\alpha} B_{\alpha} = 0
& \Longrightarrow &
B_0 = - \frac{u^i}{u^0} \, B_i = N^i B_i =
- \frac{\tfrac12 B_i B_i}{\sqrt{1 + C + \tfrac14 B_k B_k}} 
\; , \label{Bi} \\
u^{\alpha} A_{\alpha\beta} = 0
& \Longrightarrow &
A_{00} = - \frac{u^i}{u^0} \, A_{i0} = - N^i A_{i0} =
\frac{\tfrac14 A_{ij} B_i B_j}{\sqrt{1 + C + \tfrac14 B_k B_k}} 
\; , \label{A00} \\
u^{\alpha} A_{\alpha\beta} = 0
& \Longrightarrow &
A_{0i} = - N^j A_{ji} =
- \frac{\tfrac12 B_j A_{ji}}{\sqrt{1 + C + \tfrac14 B_k B_k}} 
\; . \label{A0i}
\end{eqnarray}

\subsection{Calculability}

- In the intricate subject of quantum gravity in de Sitter 
spacetime, it would appear that demanding of the argumentation 
to be supported by {\it fully regularized and renormalized 
computations} is an almost necessary, albeit highly non-trivial,
condition to meet. As expected, the requirement of complete 
calculability needs appropriate:
\\ [3pt]
{\it (i) regularization:}
We use dimensional regularization, a well-established
technique that respects general co-ordinate invariance
and includes {\it all} available modes. One merely
{\it assumes} that as the high frequency modes redshift
they equilibrate into the gravitational field equations
of general relativity.
\\ [3pt]
{\it (ii) gauge fixing condition:} It would be clearly
desirable to employ a de Sitter invariant gauge condition
but unfortunately the resulting propagator would be very
complicated. However, one could still impose an exact
de Sitter invariant gauge in which case the propagator
equation would be de Sitter invariant but would not
possess a de Sitter invariant solution.\footnote{The
classic example is the theory of a massless minimally
coupled scalar in de Sitter background
\cite{Allen:1987tz,Miao:2010vs}. Another more recent
example comes from the computation of the graviton
1-loop correction to the vacuum polarization
\cite{Glavan:2015ura}.} Moreover, the de Sitter invariant
solutions of the graviton propagator equation are
problematical because they do not correspond to proper
propagators; not {\it every} Green's function is a
propagator.\footnote{For instance, the ``propagator''
proposed in \cite{Morrison:2013rqa} although de Sitter
invariant, not only lacks calculability but also does
not correspond, as it should, to the time-ordered product
of two graviton fields \cite{Miao:2013isa}.} Provided
that physical observables are computed, it is perfectly
adequate to employ the graviton propagator that allows
regularized, BPHZ-renormalized loop calculations to be
fully completed, even if it breaks de Sitter invariance.
\footnote{This is accomplished by virtue of its
{\it coordinate independent} tensor structure and
the presence of only two ``scalar propagator'' terms
in $D \!=\! 4$.}

\subsection{The Feynman rules}

We simply display the results from \cite{Miao:2024shs}
concerning:
\footnote{Some useful properties of the propagators
and their derivatives can be found in the Appendix.}
\\ [5pt]
$\bullet$ {\it The gauge fixing sector}
\\
The extended gauge fixing Lagrangian term equals:
\begin{equation}
{\widetilde{\mathcal L}}_{GF} =
- \tfrac12 a^{D-2} \sqrt{-{\widetilde g}} \,
{\widetilde g}^{\mu\nu} \, {\widetilde F}_{\mu} \,
{\widetilde F}_{\nu}
\; , \label{Lgf1} \\
\end{equation}
and the extended gauge condition is:
\begin{equation}
{\widetilde F}_{\mu} =
{\widetilde g}^{\rho\sigma} \,
\Big[ h_{\mu\rho, \sigma} - \tfrac12 h_{\rho\sigma, \mu} \,
- (D-2) a {\widetilde H} h_{\mu\rho} u_{\sigma} \Big]
\quad , \quad
{\widetilde H} \equiv \frac{H}{N}
\; . \qquad \label{F}
\end{equation}
To arrive at the desired form for the gauge fixing
Lagrangian term, we substitute (\ref{F})
into (\ref{Lgf1}):
\begin{eqnarray}
{\widetilde{\mathcal L}}_{GF}
&\!\!\!=\!\!\!&
a^{D-2} \sqrt{-{\widetilde g}} \, {\widetilde g}^{\alpha\beta}
{\widetilde g}^{\rho\sigma} {\widetilde g}^{\mu\nu}
\Big\{
- \tfrac12 h_{\mu\rho, \sigma} h_{\nu\alpha, \beta}
+ \tfrac12 h_{\mu\rho, \sigma} h_{\alpha\beta, \nu}
- \tfrac18 h_{\rho\sigma, \mu} h_{\alpha\beta, \nu}
\qquad \nonumber \\
& \mbox{} &
+ (D-2) a {\widetilde H} h_{\mu\rho, \sigma} h_{\nu\alpha} u_{\beta}
- (\tfrac{D}{2}-1) a {\widetilde H} h_{\rho\sigma, \mu} h_{\nu\alpha}
u_{\beta}
\nonumber \\
& \mbox{} &
- \tfrac12 (D-2)^2 a^2 {\widetilde H}^2 h_{\mu\rho} h_{\nu\alpha}
u_{\beta} u_{\sigma}
\Big\}
\label{Lgf2}
\end{eqnarray}
$\bullet$ {\it The graviton sector}
\\
The graviton propagator is:
\begin{equation}
i [\mbox{}_{\mu\nu} \widetilde{\Delta}_{\rho\sigma}](x;x')
=
\sum_{I=A,B,C} [\mbox{}_{\mu\nu} \widetilde{T}^I_{~\rho\sigma}]
\!\times
i \widetilde{\Delta}_{I}(x;x')
\; , \label{gravprop}
\end{equation}
where $i \widetilde{\Delta}_{A}(x;x')$ is the massless minimally
coupled scalar propagator in dS with Hubble parameter $\widetilde{H}$,
$i \widetilde{\Delta}_{B}(x;x')$ is the massive scalar propagator in
dS for $m^2 = (D-2) \widetilde{H}^2$,
and $i \widetilde{\Delta}_{C}(x;x')$ is the massive scalar propagator
in dS for $m^2 = 2(D-3) \widetilde{H}^2$.
The tensor factors, as per usual in our class of gauges, have the
remarkable property of being spacetime {\it constants} and hence
the resulting propagator has excellent perturbative calculability:
\footnote{Furthermore, in $D=4$ the three scalar propagators
$i \widetilde{\Delta}_{I}(x;x')$ {\it only} have one or two terms.}
\begin{eqnarray}
[\mbox{}_{\mu\nu} \widetilde{T}^{A}_{~\rho\sigma}]
&\!\!\! = \!\!\!&
2 \overline{\gamma}_{\mu (\rho} \overline{\gamma}_{\sigma) \nu}
- \tfrac{2}{D-3} \overline{\gamma}_{\mu\nu} \overline{\gamma}_{\rho\sigma}
\; , \label{T_A} \\
{[} \mbox{}_{\mu\nu} \widetilde{T}^{B}_{~\rho\sigma}]
&\!\!\! = \!\!\!&
- 4 u_{(\mu} \overline{\gamma}_{\nu) (\rho} u_{\sigma)}
\; , \label{T_B} \\
{[} \mbox{}_{\mu\nu} \widetilde{T}^{C}_{~\rho\sigma}]
&\!\!\! = \!\!\!&
\tfrac{2}{(D-3)(D-2)}
\big[ (D-3) u_{\mu} u_{\nu} + \overline{\gamma}_{\mu\nu} \big]
\big[ (D-3) u_{\rho} u_{\sigma} + \overline{\gamma}_{\rho\sigma} \big]
\; . \label{T_C}
\end{eqnarray}
$\bullet$ {\it The ghost sector}
\\
The ghost and antighost fermionic fields $c_{\mu}$ and
${\overline c}_{\mu}$ contribute thusly:
\begin{equation}
{\widetilde{\mathcal L}}_{gh}
=
- a^{D-2} \sqrt{-{\widetilde g}} \, {\widetilde g}^{\mu\nu}
{\overline c}_{\mu} \, \delta F_{\nu}
\; , \label{Lgh}
\end{equation}
where in the infinitesimal variation $\delta F_{\nu}$
of the gauge fixing condition (\ref{F}) the variation
parameter is the ghost field $c$.

Perhaps a more appropriate re-writing of (\ref{Lgh}) is:
\begin{equation}
\widetilde{\mathcal{L}}_{\rm ghost}
=
a^{D-2} \sqrt{-\widetilde{g}} \, \overline{c}^{\alpha}
\, \widetilde{\mathcal{G}}_{\alpha\beta} \, c^{\beta}.
\; , \label{Lgh2}
\end{equation}
where the differential operator
$\widetilde{\mathcal{G}}_{\alpha\beta}$ equals:
\begin{eqnarray}
\widetilde{\mathcal{G}}_{\alpha\beta}
& \!\!\!=\!\!\! &
\widetilde{g}_{\alpha\beta} \widetilde{g}^{\rho\sigma}
\partial_{\rho} \partial_{\sigma}
- {\scriptstyle (D-2)} \, a \widetilde{H} \eta_{\alpha\beta} \, u^{\rho}
\partial_{\rho}
+ {\scriptstyle (D-2)} \, a^2 \widetilde{H}^2 u_{\alpha} u_{\beta}
\nonumber \\
& & \hspace{-0.9cm}
+ \, \widetilde{g}^{\rho\sigma} \kappa ( h_{\beta\rho , \sigma}
\!-\! \tfrac12 h_{\rho\sigma , \beta} ) \partial_{\alpha}
+ \big[ \, \widetilde{g}^{\rho\sigma} \kappa ( h_{\alpha\rho , \sigma}
\!-\! \tfrac12 h_{\rho\sigma , \alpha} ) \big]_{,\beta}
\nonumber \\
& & \hspace{-0.9cm}
+ {\scriptstyle (D-2)} \, a \widetilde{H} \eta_{\alpha\rho}
u_{\sigma} \widetilde{g}^{\rho\sigma}_{~~,\beta}
+ {\scriptstyle (D-2)} \, a \widetilde{H} \kappa h_{\alpha\rho}
\widetilde{g}^{\rho\sigma} u_{\beta} \partial_{\sigma}
- {\scriptstyle 2 (D-2)} \, a^2 \widetilde{H}^2 \kappa h_{\alpha\rho}
u^{\rho} u_{\beta}
\; . \qquad \label{Gtilde}
\end{eqnarray}
The ghost propagator takes the form:
\begin{equation}
i[\mbox{}_{\mu} \widetilde{\Delta}_{\rho}](x;x')
=
\overline{\gamma}_{\mu\rho} \!\times i \widetilde{\Delta}_A(x;x')
- u_{\mu} u_{\rho} \!\times i \widetilde{\Delta}_{B}(x;x')
\; . \label{ghostprop}
\end{equation}

It is important to note that relative to our standard
gauge condition \cite{Miao:2024shs}:
\begin{equation}
F_{\mu} =
\eta^{\rho\sigma} \, 
\big[ h_{\mu\rho, \sigma} - \tfrac12 h_{\rho\sigma, \mu} \,
- (D-2) \, a H h_{\mu\rho} \, \delta^0_{\sigma} \big]
\; , \label{FdS} \\
\end{equation}
adopting the extended gauge condition (\ref{F}) {\it only} 
changes the vertices; 
the graviton and ghost propagators remain the {\it same}.

\subsection{The optimal decomposition}
The total Lagrangian is the usual sum of three terms: 
the invariant part $\mathcal{L}_{inv}$, the gauge fixing
part $\mathcal{\widetilde L}_{GF}$, and the ghost part
$\mathcal{\widetilde L}_{gh}$:
\begin{equation}
\mathcal{L}_{tot} =
\mathcal{L}_{inv} + \mathcal{\widetilde L}_{GF} +
\mathcal{\widetilde L}_{gh}
\; . \label{Ltot}
\end{equation}
It will turn out that for our purposes from now onwards 
it is mathematically correct to take the $D=4$ limit.
\footnote{The integrating out procedure shall require 
the coincidence limits of the three scalar propagators 
$\widetilde{\Delta}_{I}(x;x')$ and their derivatives; 
they are given in Table \ref{Limits} of Appendix. 
An elementary inspection of Table \ref{Limits} shows 
that the {\it only} entry which is not finite is 
$i{\widetilde \Delta}_A (x;x') \vert_{x=x'}$ which 
{\it never} appears in any of the steps that will 
lead to the final form of the LLOG equations we shall 
obtain. This is the justification for computing in 
the $D=4$ limit.}

Furthermore, it is preferable to add the contributions 
of $\mathcal{L}_{inv}$ (\ref{Linv2}) and 
$\mathcal{\widetilde L}_{GF}$ (\ref{Lgf2}) because some 
of their terms simplify against each other. The six terms 
comprising $\, \mathcal{L}_{inv} \!+\! 
\mathcal{\widetilde L}_{GF} \,$ can be grouped into three 
convenient parts:
\begin{equation}
\mathcal{L}_{inv} + \mathcal{\widetilde L}_{GF}
\; \equiv \;
\mathcal{L}_{1+2+3} + \mathcal{L}_{4+5} + \mathcal{L}_6
\; , \label{Linv+Lgf}
\end{equation}
where these three convenient parts in $D=4$ equal:
\begin{eqnarray}
\mathcal{L}_{1+2+3} 
&\!\!\! = \!\!\!&
a^2 \sqrt{-{\widetilde g}} \, {\widetilde g}^{\alpha\beta}
{\widetilde g}^{\rho\sigma} {\widetilde g}^{\gamma\delta} 
\nonumber \\
& \mbox{} & \hspace{0.3cm}
\times \,\Big\{ \!
- \tfrac14 h_{\alpha\rho, \gamma} h_{\beta\sigma, \delta} 
+ \tfrac18 h_{\alpha\beta, \gamma} h_{\rho\sigma, \delta}
+ a^2 {\widetilde H}^2 h_{\gamma\rho} u_{\sigma} 
  h_{\delta\alpha} u_{\beta} \Big\}
\; , \qquad \label{L1+2+3} \\
\mathcal{L}_{4+5} 
&\!\!\! = \!\!\!&
\sqrt{-{\widetilde g}} \, {\widetilde g}^{\alpha\beta}
{\widetilde g}^{\rho\sigma} {\widetilde g}^{\gamma\delta}
\partial_{\beta} \Big[ 
- \tfrac12 \partial_{\sigma} \big( a^2 h_{\gamma\rho} 
  h_{\delta\alpha} \big)
+ a^2 h_{\gamma\rho} h_{\delta\alpha, \sigma} \Big]
\; , \label{L4+5} \\
\mathcal{L}_6
&\!\!\! = \!\!\!&
\kappa a^3 H \sqrt{-{\widetilde g}} \, {\widetilde g}^{\alpha\beta}
{\widetilde g}^{\rho\sigma} {\widetilde g}^{\gamma\delta}
h_{\rho\sigma, \gamma} h_{\delta\alpha} h_{0\beta}
\; . \label{L6} 
\end{eqnarray}
Finally the ghost contribution $\mathcal{\widetilde L}_{gh}$
of interest is given by (\ref{Lgh2}) after we set $D=4$ in 
(\ref{Gtilde}).

\subsection{The resulting equations of motion}
The starting point of our analysis are the $D=4$ gravitational
field equations that follow from $\mathcal{L}_{tot}$ when we 
consider the three convenient parts (\ref{L1+2+3}-\ref{L6}):
\begin{eqnarray}
\frac{\delta S_{1+2+3}}{\delta h_{\mu\nu}} 
& \!\!\!=\!\!\! &
\partial_{\alpha} \Bigl\{ 
a^2 \sqrt{-\widetilde{g}} \, \widetilde{g}^{\alpha\beta} 
( \tfrac12 \widetilde{g}^{\mu \rho} \widetilde{g}^{\nu \sigma} 
\!\!-\! \tfrac14 \widetilde{g}^{\mu\nu} \widetilde{g}^{\rho\sigma} ) 
h_{\rho\sigma , \beta} \Bigr\} 
\hspace{4.1cm} \nonumber \\
& &
+ 2 a^4 \sqrt{-\widetilde{g}} \, \widetilde{H}^2 
u^{(\mu} \widetilde{g}^{\nu) \alpha} u^{\beta} h_{\alpha\beta}
\nonumber \\
& & 
+ a^2 \tfrac{\partial}{\partial h_{\mu\nu}} \Bigl\{
\sqrt{-\widetilde{g}} \, \widetilde{g}^{\alpha\beta} 
( -\tfrac14 \widetilde{g}^{\rho\gamma} \widetilde{g}^{\sigma\delta} 
+ \tfrac18 \widetilde{g}^{\rho\sigma} \widetilde{g}^{\gamma\delta}) \Bigr\} 
h_{\gamma\delta , \alpha} h_{\rho\sigma , \beta} 
\nonumber \\
& & 
+ a^4 \widetilde{H}^2 \tfrac{\partial}{\partial h_{\mu\nu}} 
\Bigl\{ \sqrt{-\widetilde{g}} \, \widetilde{g}^{\gamma\alpha} 
\widetilde{g}^{\beta\rho} \widetilde{g}^{\sigma\delta} \Bigr\} 
u_{\gamma} u_{\delta} h_{\alpha\beta} h_{\rho\sigma} 
\; , \label{L1+2+3eom}
\end{eqnarray}
\begin{eqnarray}
\frac{\delta S_{4+5}}{\delta h_{\mu\nu}} 
& \!\!\!=\!\!\! & 
\tfrac{\partial}{\partial h_{\mu\nu}} \Bigl\{ 
\sqrt{-\widetilde{g}} \, \widetilde{g}^{\alpha\beta} 
\widetilde{g}^{\gamma\delta} \widetilde{g}^{\rho\sigma} \Bigr\}
\hspace{7.7cm} \nonumber \\
& & \hspace{1.3cm}
\times \Bigl\{ -3 a^4 \widetilde{H}^2 u_{\beta} u_{\sigma} 
h_{\gamma\rho} h_{\delta\alpha} 
+ \tfrac{a^2}{2} ( h_{\gamma\rho , \beta} h_{\delta\alpha , \sigma}  
- h_{\gamma\rho , \sigma} h_{\delta\alpha , \beta} ) \Bigr\} 
\nonumber \\
& &
+ a^3 \sqrt{-\widetilde{g}} \, \widetilde{H} 
( 2 \widetilde{g}^{\alpha(\mu} u^{\nu)} 
\!\!-\! u^{\alpha} \widetilde{g}^{\mu\nu} )
h_{\alpha\beta} \, \widetilde{g}^{\beta \rho}_{~~,\rho} 
\nonumber \\
& & 
- 2 a^3 (\widetilde{H} \sqrt{-\widetilde{g}} \, u^{\alpha})_{,\rho} 
\, h_{\alpha\beta} \, \widetilde{g}^{\beta (\mu} 
\widetilde{g}^{\nu) \rho} 
\nonumber \\
& &
+ a^2( h_{\gamma\rho , \sigma} \partial_{\beta} 
\!-\! h_{\gamma\rho , \beta} \partial_{\sigma} ) 
(\sqrt{-\widetilde{g}} \, \widetilde{g}^{\rho\sigma}
\widetilde{g}^{\beta (\mu} \widetilde{g}^{\nu ) \gamma}) 
\; , \label{L4+5eom}
\end{eqnarray}

\newpage

\begin{eqnarray}
\frac{\delta S_{6}}{\delta h_{\mu\nu}}  
& \!\!\!=\!\!\! & 
-\kappa H \sqrt{-\widetilde{g}} \, \widetilde{g}^{\mu\nu} 
\partial_{\gamma} ( a^3 \widetilde{g}^{\alpha\beta} 
\widetilde{g}^{\gamma\delta} h_{\delta \alpha} h_{\beta 0} )
\hspace{5.9cm} \nonumber \\
& &
+ 2 a^3 H \tfrac{\partial}{\partial h_{\mu\nu}} \Bigl\{
\widetilde{g}^{\alpha\beta} \widetilde{g}^{\gamma\delta} 
h_{\alpha\delta} h_{\beta 0} \Bigr\} 
\partial_{\gamma} \sqrt{-\widetilde{g}} 
\; , \label{L6eom}
\end{eqnarray}
as well as the $D=4$ limit of the ghost Lagrangian (\ref{Lgh2})
that gives the most complicated equation of motion by far: 
\begin{eqnarray}
\frac{\delta \widetilde{S}_{\rm ghost}}{\delta h_{\mu\nu}} 
& \!\!\!=\!\!\! & 
\tfrac{\kappa}{2} a^2 \sqrt{-\widetilde{g}} \, \widetilde{g}^{\mu\nu} \, 
\overline{c}^{\alpha} \widetilde{\mathcal{D}}_{\alpha\beta} \, c^{\beta} 
+ \kappa a^2 \sqrt{-\widetilde{g}} \, \overline{c}^{\alpha} \Bigl\{ 
( \delta^{\mu}_{\alpha} \delta^{\nu}_{\beta} \, \widetilde{g}^{\rho\sigma} 
\!-\! \widetilde{g}_{\alpha\beta} \widetilde{g}^{\rho\mu} 
\widetilde{g}^{\sigma\nu} ) \partial_{\rho} \partial_{\sigma}
\nonumber \\
& & \hspace{-0.7cm}
+ 2 a \widetilde{H} \eta_{\alpha\beta} \, u^{(\mu} \widetilde{g}^{\nu) \rho} 
\partial_{\rho} 
\!-\! \widetilde{g}^{\rho\mu} \widetilde{g}^{\sigma\nu} \kappa 
( h_{\beta \rho , \sigma} \!-\! \tfrac12 h_{\rho\sigma , \beta} ) 
\partial_{\alpha} 
\!+\! 2 a \widetilde{H} \eta_{\alpha\rho} \, \widetilde{g}^{\rho (\mu} 
\widetilde{g}^{\nu) \sigma} ) u_{\beta} \partial_{\sigma} 
\nonumber \\
& & 
- 4 a^2 \! \widetilde{H}^2 \eta_{\alpha\rho} \,
\widetilde{g}^{\rho (\mu} u^{\nu)} u_{\beta} \Bigr\} \, c^{\beta}
\nonumber \\
& & \hspace{-0.7cm}
- \kappa \, \partial_{\rho} \Bigl\{a^2 \sqrt{-\widetilde{g}} \, 
\overline{c}^{\alpha} \widetilde{g}^{\rho (\mu} c^{\nu)}_{~~,\alpha} \Bigr\} 
+ \tfrac{\kappa}{2} \, \partial_{\beta} \Bigl\{ 
a^2 \sqrt{-\widetilde{g}} \, \widetilde{g}^{\mu\nu} \, 
\overline{c}^{\alpha} c^{\beta}_{~,\alpha} \Bigr\} 
\nonumber \\
& & \hspace{-0.7cm}  
+ \, \widetilde{g}^{\rho (\mu} \widetilde{g}^{\nu) \sigma} 
\kappa^2 ( h_{\alpha\rho , \sigma} - \tfrac12 h_{\rho\sigma , \alpha} ) \,
\partial_{\beta} \Bigl\{ a^2 \sqrt{-\widetilde{g}} \, 
\overline{c}^{\alpha} c^{\beta} \Bigr\}
\nonumber \\
& & \hspace{-0.7cm}
+ \kappa \, \partial_{\rho} \Bigl\{ 
\widetilde{g}^{\rho (\mu} \partial_{\beta} 
( a^2 \sqrt{-\widetilde{g}} \, \overline{c}^{\nu)} c^{\beta} ) \Bigr\}
- \tfrac{\kappa}{2} \, \partial_{\alpha} \Bigl\{ 
\widetilde{g}^{\mu\nu} \partial_{\beta} ( a^2 \sqrt{-\widetilde{g}} \, 
\overline{c}^{\alpha} c^{\beta}) \Bigr\} 
\nonumber \\
& & \hspace{-0.7cm}
+ 2 \, \widetilde{g}^{\rho (\mu} \widetilde{g}^{\nu) \sigma} \partial_{\beta}
\Bigl\{ a^3 \sqrt{-\widetilde{g}} \, \kappa \widetilde{H} \eta_{\alpha\rho} 
u_{\sigma} \, \overline{c}^{\alpha} c^{\beta} \Bigr\} 
\; , \label{Lgheom}
\end{eqnarray}
where $\widetilde{\mathcal{D}}_{\alpha\beta}
= \widetilde{\mathcal{G}}_{\alpha\beta}$ in $D=4$.

\section{LLOG Approximation Equations}

Dimensionally regularized, and fully renormalized perturbative
computations can exhibit secular effects characterized by 
powers of $ln(a)$ due to two sources:
\\ [3pt]
{\bf (i)} The logarithmic part of the propagator 
{\it (stochastic logs)} which can be schematically
expressed as:
\footnote{DeWitt and Brehme termed this part of the graviton
propagator as the ``tail'' part \cite{DeWitt:1960fc}.}
\begin{equation}
i \Delta \sim \tfrac{1}{x^2} - \ln(x^2)
\; , \label{TailLogs}
\end{equation}
Quantum fields which possess such ``tails'' are known as 
{\it active}; all other fields are termed {\it passive}.
\\ [3pt]
{\bf (ii)} The incomplete cancellation between primitive 
divergences and counterterms {\it (renormalization induced 
logs)} which can be schematically expressed as:
\begin{equation}
\tfrac{H^{D-4}}{D-4} - \tfrac{(\mu a)^{D-4}}{D-4}
=
- \ln\left( \tfrac{\mu a}{H} \right) + \mathcal{O}(D-4)
\; . \label{UVlogs}
\end{equation}
Any type of quantum field can contribute this kind of 
logarithm.

$\ast$ {\it Case I:} When interactions contain only 
undifferentiated active fields, the leading logarithms 
derive entirely from (\ref{TailLogs}) and can be recovered 
by replacing the full quantum field with a stochastic 
random field described by Starobinsky's formalism
\cite{Starobinsky:1986fx,Starobinsky:1994bd}. 
\footnote{The amazing fact that (ultraviolet divergent) 
correlators of the full quantum field $\phi(x)$ agree, 
at leading logarithm order, with those of the stochastic 
random field was proved in \cite{Tsamis:2005hd}.}

$\ast$ {\it Case II:} When undifferentiated active fields 
interact with passive fields, the leading logarithms again 
derive from the tail (\ref{TailLogs}). However, it is not 
correct to stochastically truncate passive fields because 
their effects derive as much from the ultraviolet as from 
the infrared, and from their full mode functions. The leading 
logarithms are instead recovered by integrating out the 
passive fields in a constant active field background, which 
produces a standard effective potential through the dependence 
of the passive field mass on the active field. One then
applies Starobinsky's formalism to the effective potential. 
\footnote{Examples include a fermion Yukawa-coupled to a 
massless minimally coupled scalar \cite{Miao:2006pn} and 
massless minimally couled  scalar electrodynamics 
\cite{Prokopec:2007ak, Prokopec:2008gw}.}

$\ast$ {\it Case III:} When interactions involve 
differentiated active fields, one generally encounters both 
kinds of logarithms.\footnote{Examples include non-linear 
sigma models \cite{Miao:2021gic}, scalar corrections to gravity 
\cite{Miao:2024nsz,Miao:2024atw} and pure gravity 
\cite{Miao:2024shs,Miao:2025gzm}.}
Stochastic logarithms (\ref{TailLogs}) are captured by integrating 
out the differentiated fields in the presence of a constant active 
field background. This produces new kinds of effective potentials 
based on how the active background affects field strengths or 
geometries. Renormalization induced logarithms (\ref{UVlogs}) 
can be treated using a variant of the renormalization group by 
exploiting the close relation between the scale factor $a(t)$
and the regularization scale $\mu$, which is evident in 
expression (\ref{UVlogs}). The variation consists of expressing 
Bogoliubov-Parasiuk-Hepp-Zimmermann (BPHZ) counterterms as a 
higher derivative of the quantum field, plus renormalizations 
of bare parameters, for example:
\begin{equation}
R^2 = (R - D \Lambda)^2 + 2 D \Lambda [R - (D\!-\!2) \Lambda)]
+ D (D \!-\! 4) \Lambda^2 \; . \label{Eddington}
\end{equation}
While the factors of $\ln(a)$ from the first term are suppressed 
by powers of $a$, those from the second term can be summed using 
the Callan-Symanzik equation \cite{Miao:2024nsz}.

In a long period of inflation, the factors of $\ln[a(t)]$ can 
grow large enough to overwhelm even the smallest coupling constant. 
The dimensionless coupling constant of pure quantum gravity is 
$G H^2$ and at some time the secular increase by powers of 
$\ln[a(t)]$ will overwhelm $G H^2$ causing perturbation theory 
to break. 

During the inflationary epoch, the various species of elementary 
particles respond to these quantum effects according to their 
interaction properties:
\\ [3pt]
{\bf -} Fermions, photons, conformal massless scalars can
only provide RG effects since none are produced due to their
conformally invariant interactions: ${\rm N}(t, k) = 0$.
\\ [3pt]
{\bf -} Massless minimally coupled scalars are produced with
${\rm N}(t, k) \! = \! \big[ \tfrac{H a(t)}{2k} \big]^2$ but 
can only provide RG effects because their only coupling to 
gravity is very weak due to the de Sitter redshift and thus 
cannot support stochastic effects.
\\ [3pt]
{\bf -} Gravitons are also produced with
${\rm N}(t, k) \! = \! \big[ \tfrac{H a(t)}{2k} \big]^2$ and
can provide both stochastic and RG effects.

Physically, the most interesting particle to study is clearly 
the graviton; can the universally attractive nature of the 
gravitational interaction alter cosmological parameters, 
kinematical parameters and long-range forces? A preliminary 
study of this, albeit with ``semi-primitive'' for the intended 
purpose quantum field theoretic tools, {\it indicated} 
a positive answer \cite{Tsamis:1996qm,Tsamis:2011ep}.

\subsection{The generic LLOG equation and rules}

The rules for obtaining the correct LLOG equations in ascending 
in complication quantum field theories for which the equations 
could be obtained and checked, were presented in \cite{Miao:2024shs, 
Miao:2025gzm}. Therein, the straightforward extension of the scalar 
rules to tensorial pure quantum gravity was suggested. Furthermore, 
it was mentioned that perturbative consistency checks were necessary 
to solidify their correctness. 

The fundamental LLOG equation given in \cite{Miao:2025gzm} is:
\begin{equation}
\frac{\delta S[h]}{\delta h_{\mu\nu}} \Big\vert_{LLOG}
\equiv
\frac{\delta S[h]}{\delta h_{\mu\nu}} \Big\vert_{stoch}
\! +
\frac{\delta S[h]}{\delta h_{\mu\nu}} \Big\vert_{ind}
\; \big( \! +
\frac{\delta S_{matter}}{\delta h_{\mu\nu}} \big)
 = 0
\; . \label{QGrule}
\end{equation}

\newpage

In the above expression (\ref{QGrule}):
\\ [3pt]
{\bf -} $\tfrac{\delta S[h]}{\delta h_{\mu\nu}} \vert_{LLOG}$
represents the desired equation of motion capturing the leading
logarithms to all orders, and is derived from the full Heisenberg
equation of motion $\tfrac{\delta S[h]}{\delta h_{\mu\nu}}$ by
adding the three contributions of the RHS.
\\ [3pt]
{\bf -} The first contribution
$\tfrac{\delta S[h]}{\delta h{\mu\nu}} \vert_{stoch}$
represents the {\it ``stochastic reduction''} of the derivative
terms in the Heisenberg field equation
$\tfrac{\delta S[h]}{\delta h_{\mu\nu}}$.
\footnote{An enormous advantage of the method becomes
apparent because we must now deal with a classical 
stochastic equation instead of the Heisenberg field 
equations of an interacting QFT.}
\\ [3pt]
{\bf -} The second contribution
$\tfrac{\delta S[h]}{\delta h{\mu\nu}} \vert_{ind}$
represents the quantum induced stress tensor resulting
from {\it ``integrating out''} the differentiated
gravitons from the equations of motion in the presence
of a constant graviton background.
\\ [3pt]
{\bf -} The third contribution
$\tfrac{\delta S_{matter}}{\delta h{\mu\nu}}$ is present
in the most general case and represents classical matter
sources.
\\ [3pt]
\noindent A very schematical way to think of the LLOG equation
(\ref{QGrule}) is that it consists of a ``kinetic'' part 
$\tfrac{\delta S[h]}{\delta h{\mu\nu}} \vert_{stoch}$ which 
is stochastic and a ``potential'' part
$\tfrac{\delta S[h]}{\delta h{\mu\nu}} \vert_{ind}$ which
is the induced stress tensor.
\\ [7pt]
\hspace{0.3cm} $\ast$ {\it The ``stochastic reduction'' rules} 
associated with (\ref{QGrule}) and given in \cite{Miao:2025gzm} 
are:
\\ [3pt]
{\bf I.} At each order in the dynamical $A_{\rho\sigma}$ 
field retain only the terms with  no derivatives {\it and} 
with the smallest number of derivatives.
\footnote{For instance, products of differentiated
$A_{\rho\sigma}$ fields are ignored.}
\\ [3pt]
{\bf II.} For the linear terms in $A_{\rho\sigma}$, 
each time derivative has a stochastic source
$\mathcal{A}_{\rho\sigma}$ subtracted.
\\ [3pt]
{\bf III.} All non-linear terms in the constrained 
fields $B_{\sigma}$ and $C$ are ignored. The same 
is true for terms which involve the product of a 
constrained field times a differentiated dynamical 
field $A_{\rho\sigma}$.
\footnote{The constrained fields $B_{\sigma}$ and
$C$, as expected, are driven by the dynamical field
$A_{\rho\sigma}$ and vanish when the latter field
vanishes. Because $A_{\rho\sigma}$ evolves slowly we
have schematically $B \!\sim\! C \!\sim\! \partial A
\!\sim\! small$, so that any products of them are even
smaller and can be ignored.}
\\ [7pt]
$\ast$ {\it The ``integrating out'' rules} associated
with (\ref{QGrule}) and given in \cite{Miao:2025gzm} are:
\\ [3pt]
{\bf IV.} Integrate out the differentiated graviton fields
from the effective gravitational equations 
(\ref{L1+2+3eom}-\ref{L6eom}).
\footnote{The gravitational interactions possess up to two 
derivatives so that at most graviton field bilinears will 
possess derivatives. The process of integrating out these 
singly or doubly differentiated graviton bilinears amounts 
to replacing them with singly or doubly differentiated 
graviton propagators in the presence of a constant graviton 
background, and this is tantamount to replacing them with 
singly or doubly differentiated de Sitter graviton propagators 
with a different Hubble parameter.}
\\ [3pt]
{\bf V.} Integrate out the ghost fields from the ghost
equation (\ref{Lgheom}).
\footnote{The ghost equation (\ref{Lgheom}) also
contains differentiated graviton fields. If we also
integrate out the latter, we shall produce terms
with the same logarithmic content but their overall
coefficient would then be $\kappa^4 {\widetilde H}^4$
and hence sub-leading with respect to the
$\kappa^2 {\widetilde H}^2$ coefficient produced 
by only integrating out the ghost fields.}

\subsection{The problem \& its solution}

Unfortunately, the LLOG rules I-V turned out to be inadequate
to accomplish agreement upon checking their results against 
those of exact perturbative computations like the Newtonian
potential due to a point mass and the graviton 1-loop 1PI 
1-point function.

It should be apparent that the reason for this inadequacy
is a property of gravitation within its basic structure that 
is not shared by the QFT's from which rules I-V emanated. 
Gravitation contains {\it constrained} degrees of freedom 
that the QFT's did not. And it turned out that the constrained 
degrees of freedom of gravity were not properly treated by 
rules IV-V. 

As we shall see, agreement with the aforementioned two 
perturbative checks requires the following addition to 
the rules:
\\ [3pt]
{\bf VI.} Integrate out the undifferentiated constrained fields
$B_{\sigma}$ and $C$ from the effective gravitational equations 
(\ref{L1+2+3eom}-\ref{Lgheom}).

We proceed to reduce equations (\ref{L1+2+3eom}-\ref{Lgheom}) 
according to the rules I-VI.

\subsection{The stochastic reduction}

Applying the reduction rules I-III to the gravitational 
equations (\ref{L1+2+3eom}-\ref{L6eom}) gives: 
\begin{eqnarray}
\kappa \,
\frac{\delta S[h]_{(1+2+3)}}{\delta h_{\mu\nu}} \Big\vert_{stoch}
& \!\!\!=\!\!\! &
\big[ {\overline \gamma}^{\mu k} \, {\overline \gamma}^{\nu l}
- {\overline \gamma}^{\mu\nu} \, {\overline \gamma}^{kl} \big]
\Big\{ \tfrac32 \, a^4 \sqrt{-\widetilde{g}} N^{-2} H
\big( {\dot A}_{kl} - {\dot {\mathcal A}}_{kl} \big) \Big\}
\nonumber \\
& &
+ \, \tfrac12 u^{(\mu}{\overline \gamma}^{\nu)k}
\Big\{ {\widetilde D}_B B_k + 2 {\widetilde H}^2 \sqrt{-\widetilde{g}}
\, A_{kl} \, {\overline \gamma}^{lm} B_m \Big\}
\nonumber \\
& &
+ \, u^{\mu} u^{\nu} \Big\{ {\widetilde D}_C C - \tfrac34 \, N^{-2} H
a^4 \sqrt{-\widetilde{g}} \, {\overline \gamma}^{kl}
\big( {\dot A}_{kl} - {\dot {\mathcal A}}_{kl} \big) \Big\}
\; , \qquad \label{S_1+2+3_stoch}
\end{eqnarray}
\begin{eqnarray}
\kappa \,
\frac{\delta S[h]_{(4+5)}}{\delta h_{\mu\nu}} \Big\vert_{stoch}
& \!\!\!=\!\!\! &
a^4 \sqrt{-\widetilde{g}} \, u^{(\mu} \, {\overline \gamma}^{\nu)i}
A_{ij} \, {\overline \gamma}^{jk} \Big[ \! 
- 3 \widetilde{H}^2 B_k - \tfrac{2}{a} \, {\widetilde H} \,
\partial_k C \Big]
\nonumber \\
& &
+ \, a^4 \sqrt{-\widetilde{g}} \Big[ \!
- \tfrac{1}{a} \, {\widetilde H} \, {\overline \gamma}^{\mu i} \,
{\overline \gamma}^{\nu j} A_{ik} \, {\overline \gamma}^{kl} \,
\partial_j B_l \Big]
\; , \label{S_4+5_stoch}
\end{eqnarray}
\begin{eqnarray}
\kappa \,
\frac{\delta S[h]_{(6)}}{\delta h_{\mu\nu}} \Big\vert_{stoch}
& \!\!\!=\!\!\! &
a^3 \sqrt{-\widetilde{g}} \, \Bigl\{
\tfrac12 \big[ {\overline \gamma}^{\mu\nu} \! - u^{\mu} u^{\nu} \big] 
N H {\gamma}^{ij} {\gamma}^{kl} A_{ik} \, \partial_j B_l
\nonumber \\
& &
+ \, 2 H \delta^{(\mu}_{~~0} \, {\overline \gamma}^{\nu)i} \,
{\overline \gamma}^{j\kappa} A_{ij} \, \partial_{\kappa} C \Bigr\}
\; . \label{S_6_stoch}
\end{eqnarray}

We then add contributions (\ref{S_1+2+3_stoch}-\ref{S_6_stoch})
and multiply their sum with the tensor projectors 
${\widetilde g}_{\mu i} \, {\widetilde g}_{\nu j}$, 
$-4 u_{\mu} \, {\widetilde g}_{\nu i}$, 
and $u_{\mu} u_{\nu}$ to obtain separate equations for the fields
$A_{ij}$, $B_i$, and $C$ respectively:
\begin{eqnarray}
\kappa \,
\frac{\delta S[h]_{A}}{\delta h_{\mu\nu}} \Big\vert_{stoch}
& \!\!\!=\!\!\! &
- \tfrac32 \, a^4 \sqrt{-{\widetilde g}} \, N^{-2} H  
\big( {\dot A}_{ij} - {\dot {\mathcal A}}_{ij} \big)
- a^4 \sqrt{-{\widetilde g}} \, {\widetilde H} A_{ik} \gamma^{kl} 
\, \tfrac{1}{a} \, \partial_j B_l
\qquad \nonumber \\
& &
- 2a^4 \sqrt{-{\widetilde g}} \, H 
\big[ \gamma_{ij} A_{kl} - \gamma_{ik} A_{jl} \big]
N^k \gamma^{lm} \, \tfrac{1}{a} \, \partial_m C
\; , \label{S_stoch_A}    
\end{eqnarray}
\begin{eqnarray}
\kappa \,
\frac{\delta S[h]_{B}}{\delta h_{\mu\nu}} \Big\vert_{stoch}
& \!\!\!=\!\!\! &
{\widetilde D}_B B_i 
- 2 {\widetilde H}^2 a^4 \sqrt{-\widetilde{g}} \, A_{ij} \gamma^{jk} B_k 
\nonumber \\
& &
+ \, a^4 \sqrt{-\widetilde{g}} \, 4 {\widetilde H} (N^2 -1) A_{ij} \gamma^{jk} 
\, \tfrac{1}{a} \, \partial_k C
\; , \label{S_stoch_B}    
\end{eqnarray}
\begin{eqnarray}
\kappa \,
\frac{\delta S[h]_{C}}{\delta h_{\mu\nu}} \Big\vert_{stoch}
& \!\!\!=\!\!\! &
{\widetilde D}_C C 
- \tfrac34 \, N^{-2} H a^4 \sqrt{-\widetilde{g}} \, {\overline \gamma}^{ij}
\big( {\dot A}_{kl} - {\dot {\mathcal A}}_{kl} \big)
\nonumber \\
& &
+ \, a^4 \sqrt{-\widetilde{g}} \, \tfrac12 \, N H \gamma^{ij} \gamma^{kl}
A_{ik} \, \tfrac{1}{a} \, \partial_j B_l
\; . \label{S_stoch_C}    
\end{eqnarray}

\subsection{The induced stress tensor }

Applying the reduction rules IV-VI to the gravitational
equations (\ref{L1+2+3eom}-\ref{L6eom}) gives:
\begin{eqnarray}
\kappa \,
\frac{\delta S[h]_{(1+2+3)}}{\delta h_{\mu\nu}} \Big\vert_{ind}
& \!\!\!=\!\!\! &
a^4 \sqrt{-{\widetilde g}} \, \tfrac{\kappa^2 {\widetilde H}^4}{8\pi^2}
\Big\{ \! -\tfrac32 \, {\overline \gamma}^{\mu\nu} + \tfrac{19}{2} \,
u^{\mu} u^{\nu} 
\nonumber \\
& &
+ \big[ \big( \! -\tfrac34 \, {\overline \gamma}^{\mu\nu}
-\tfrac{17}{4} \, u^{\mu} u^{\nu} \big) \gamma^{ij}
+ \tfrac92 \, {\overline \gamma}^{\mu i} \, {\overline \gamma}^{\nu j}
\big] A_{ij}
\nonumber \\
& &
+ \big[ \big( \tfrac12 \, {\overline \gamma}^{\mu\nu}
+\tfrac32 \, u^{\mu} u^{\nu} \big) \gamma^{ik} \gamma^{jl}
-3 {\overline \gamma}^{\mu i} \gamma^{jk} \, {\overline \gamma}^{l \nu} 
\big] A_{ij} A_{kl} \Big\} 
\; , \qquad \label{S_1+2+3_Tmn}
\end{eqnarray}
\begin{eqnarray}
\kappa \,
\frac{\delta S[h]_{(4+5)}}{\delta h_{\mu\nu}} \Big\vert_{ind}
& \!\!\!=\!\!\! &
a^4 \sqrt{-{\widetilde g}} \, \tfrac{\kappa^2 {\widetilde H}^4}{8\pi^2}
\Big\{ \! - 3 u^{\mu} u^{\nu} 
\nonumber \\
& &
+ \big[ \big( 3 {\overline \gamma}^{\mu\nu}
+ 6 u^{\mu} u^{\nu} \big) \gamma^{ij} \!
- 12 {\overline \gamma}^{\mu i} \, {\overline \gamma}^{\nu j}
\big] A_{ij}
\nonumber \\
& &
+ \big[ \big( \! -\tfrac32 \, {\overline \gamma}^{\mu\nu} \!
- \tfrac92 \, u^{\mu} u^{\nu} \big) \gamma^{ik} \gamma^{jl}
+ 9 {\overline \gamma}^{\mu i} \gamma^{jk} \, {\overline \gamma}^{l \nu} 
\big] A_{ij} A_{kl} \Big\} 
\; , \qquad \label{S_4+5_Tmn}
\end{eqnarray}
\begin{eqnarray}
\kappa \,
\frac{\delta S[h]_{(6)}}{\delta h_{\mu\nu}} \Big\vert_{ind}
& \!\!\!=\!\!\! &
a^4 \sqrt{-{\widetilde g}} \, \tfrac{\kappa^2 H^2 {\widetilde H}^2}{8\pi^2}
\Big\{ 3 {\overline \gamma}^{\mu\nu} \! - 3 u^{\mu} u^{\nu} 
+ 3 \big( {\overline \gamma}^{\mu\nu} \! - u^{\mu} u^{\nu} 
\big) \gamma^{ij} A_{ij} \Big\}
\; , \qquad \label{S_6_Tmn}
\end{eqnarray}
\begin{eqnarray}
\kappa \,
\frac{\delta S[h]_{gh}}{\delta h_{\mu\nu}} \Big\vert_{ind}
& \!\!\!=\!\!\! &
a^4 \sqrt{-{\widetilde g}} \, \tfrac{\kappa^2 {\widetilde H}^4}{8\pi^2}
\Big\{ \! - \tfrac32 \, {\overline \gamma}^{\mu\nu} \!
- \tfrac52 \, u^{\mu} u^{\nu}
+ \big( {\overline \gamma}^{\mu\nu} \! + u^{\mu} u^{\nu} 
\big) \gamma^{ij} A_{ij} \Big\}
\; . \qquad \label{S_gh_Tmn}
\end{eqnarray}
We can add the contributions (\ref{S_1+2+3_stoch}-\ref{S_6_stoch})
to get the total induced stress tensor:
\begin{eqnarray}
\kappa \,
\frac{\delta S[h]_{tot}}{\delta h_{\mu\nu}} \Big\vert_{ind}
& \!\!\!=\!\!\! &
a^4 \sqrt{-{\widetilde g}} \, \tfrac{\kappa^2 {\widetilde H}^4}{8\pi^2}
\Big\{ u^{\mu} u^{\nu}
+ \big[ \big( \tfrac{25}{4} \, {\overline \gamma}^{\mu\nu} \!
- \tfrac14 \, u^{\mu} u^{\nu} \big) \gamma^{ij}
- \tfrac{15}{2} \, {\overline \gamma}^{\mu i} \, {\overline \gamma}^{\nu j}
\big] A_{ij}
\nonumber \\
& &
+ \big[ \big( \! - {\overline \gamma}^{\mu\nu} \!
- 3 u^{\mu} u^{\nu} \big) \gamma^{ik} \gamma^{jl}
+ 6 {\overline \gamma}^{\mu i} \gamma^{jk} \, {\overline \gamma}^{l \nu} 
\big] A_{ij} A_{kl} \Big\} 
\; , \qquad \label{S_tot_Tmn}
\end{eqnarray}
and then obtain separate equations for the fields $A_{ij}$, $B_i$, 
and $C$, by multiplying their sum (\ref{S_tot_Tmn}) with 
${\widetilde g}_{\mu i} \,
{\widetilde g}_{\nu j}$, $-4 u_{\mu} \, {\widetilde g}_{\nu i}$,
and $u_{\mu} u_{\nu}$ respectively.


\subsection{The updated form of the LLOG equations}

Combining the results of subsections 3.3 \& 3.4 and substituting 
them in the fundamental form (\ref{QGrule}):
\begin{equation} 
\kappa \frac{\delta S[h]_I}{\delta h_{\mu\nu}} \Big\vert_{stoch}
= 
- \kappa \frac{\delta S[h]_I}{\delta h_{\mu\nu}} \Big\vert_{ind}
\quad , \quad I = A, B, C
\; , \label{QGrule2}
\end{equation}
gives the desired LLOG equations:
\footnote{Besides rules I-III \& rules IV-VI, identities from 
the Appendix and have been used for further simplifications.}
\begin{eqnarray}
& &
3 H \big( \dot{A}_{ij} - \dot{\mathcal A}_{ij} \big)
\; = \;
\tfrac{\kappa^2 H^4}{8 \pi^2} \Big\{
\! - \! \big( 6 \, \delta^k_i \delta^l_j + 16 \gamma_{ij} \, \gamma^{kl} 
\big) A_{kl}
\nonumber \\
& & 
\hspace{3.3cm}
+ \big( 12 \, \delta^k_i \delta^n_j - 8 \gamma_{ij} \, \gamma^{kn} 
\big) \gamma^{lm} A_{kl} A_{mn} \Big\}
\; , \label{A_final} \\
& &
{\widetilde D}_B B_i 
\; = \;
0
\; , \label{B_final} \\
& &
{\widetilde D}_C C 
\; = \;
\tfrac{\kappa^2 {\widetilde H}^4}{8 \pi^2} a^4 \sqrt{\gamma} \Big\{
\! - 1 - 14 \gamma^{kl} A_{kl} \Big\}
\; . \label{C_final}
\end{eqnarray}
The set of operator LLOG equations (\ref{A_final}-\ref{C_final})
contain the leading logarithms from all orders of perturbation 
theory and can be seen to be a {\it huge} simplification of the 
original field equations of pure gravity (\ref{eom}) in the gauge 
condition (\ref{F}).
\footnote{It is worth mentioning at this stage that classifying 
according to the strength of the logarithmic terms is a 
well-defined procedure. Because pure gravity is a two-parameter 
theory -- $G$ \& $\Lambda$ -- and the dimensionless coupling 
constant is the product $G \Lambda \, \ln(a)$, we can classify 
diagrams in a well-defined way by $G \Lambda$ or by $\ln(a)$.}

The operator equations (\ref{A_final}-\ref{C_final}), 
have the ability to obtain the complete (non-perturbative) 
LLOG time evolution of a physical quantity; this is still 
a highly non-trivial project to say the least but now it 
is a project which has perhaps become a rather realistic 
endeavour. Moreover, there is another way to use this set
of equations by considering their perturbative order by 
order limit; doing so permits us to easily determine the 
leading logarithm result of high-loop calculations which 
would normally take a long time to conclude.

More importantly, to procure {\it gauge independent} 
results from the set of gauge dependent {\it operator} 
equations (\ref{A_final}-\ref{C_final}), we can take 
the vacuum expectation value of a {\it gauge invariant} 
operator.

\newpage

\section{LLOG eqs: The Newtonian potential check}

Computing the Newtonian potential from the revised 
LLOG equations offers a relatively simple non-trivial
check on their correctness notably because the relevant 
gravitational equation comes from the constrained 
sector of gravitation which is precisely the sector 
whose LLOG rules we altered. Moreover, the first 
non-trivial corrections to the Newtonian potential 
have been already computed in perturbation theory 
\cite{Tan:2022xpn}: 
\footnote{In \cite{Tan:2022xpn} the Newtonian potential 
is denoted by $\Psi$ and versus definition (\ref{force}) 
the overall constant is $\, C = -2 \Psi$.}
\begin{equation}
\Psi \longrightarrow
\frac{G M}{a r} \Big\{ 1 + \frac{8 G H^2}{\pi} \big[
- \ln^3 a + \dots \big] + \dots \Big\}
\; , \label{tanresult}
\end{equation}	  
making the comparison with the LLOG result rather 
simple.

The gravitational force $F$ is defined in terms of the 
constrained gravitational variable $C$ (\ref{ABC}) in 
the usual way:
\begin{equation}
F(t, r) = - \nabla C(t, r) 
\; . \label{force}
\end{equation}
The relevant gravitational equation appropriate to give 
the response to the presence of a point mass $M$ comes 
from the constraint field equation involving $C$:
\footnote{By {\it (subdominant)} in (\ref{Ceqn}) we 
mean terms of (\ref{C_final}) which cannot possibly 
give the cubic dependence on $\ln a$ that the perturbative 
calculation (\ref{tanresult}) contains \cite{Miao:2026cfl}.}
\begin{equation}
\widetilde{D}_C \, C \, = \, 
8 \pi G M a \, \delta^3({\bf x}) + (subdominant) 
\; , \label{Ceqn}
\end{equation}
where the differential operator $\widetilde{D}_C$ is given 
by \cite{Miao:2025gzm}:
\footnote{The form of the three graviton quadratic operators
$\widetilde{D}_I, \; I={A,B,C}$ can be found in the Appendix.}
\begin{equation}
\widetilde{D}_C \equiv
\partial_{\mu} \Bigl[ a^2 \sqrt{-\widetilde{g}} 
   \, \widetilde{g}^{\mu\nu} \partial_{\nu} \Bigr] 
- 2 \widetilde{H}^2 a^4 \sqrt{-\widetilde{g}} 
\; . \label{DC}
\end{equation}
It turns out that in the first part of the right
hand side of (\ref{DC}), terms of ${\mathcal O}(\kappa)$ 
and higher cannot furnish the terms proportional to 
$\ln^3 a$ as agreement with (\ref{tanresult}) dictates 
\cite{Miao:2026cfl}. Thus, we can reduce (\ref{DC}) to:
\begin{equation}
\partial^{\mu} \big[ a^2 \partial_{\mu} C \big] 
- 2 H^2 a^4 C - 2 H^2 a^4 \tfrac{\kappa}{2} h \,C
\simeq
8 \pi G M a \, \delta^3({\bf x})
\; , \label{Ceqn2}
\end{equation}
and by going to co-moving time $t$ we reach the
following form of the equation:
\begin{equation}
- \big[ \partial^2_t + 3 H \partial_t + 2 H^2 - \tfrac{\nabla^2}{a^2} \big]C
\simeq
\kappa H^2 h \, C + 8 \pi G M \, \tfrac{\delta^3({\bf x})}{a^3}
\; , \label{Ceqn3}
\end{equation}
which we shall solve order by order:
\begin{equation}
C(t,{\bf x}) \equiv 
C^{(0)}(t,{\bf x}) + C^{(1)}(t,{\bf x}) + C^{(2)}(t,{\bf x}) + \dots
\; . \label{Csol}
\end{equation}
\\
$\bullet \; 0^{th}order \; solution.$ It equals the classical 
Newtonian potential:
\footnote{In (\ref{Ceqnsol0}) and thereafter, the
$\tfrac{\nabla^2}{a^2}$ term is sub-dominant with
respect to time derivatives and also suppressed due
to the presence of inverse scale factor powers;
furthermore, the remaining differential operator
equals $\tfrac{1}{a^2} \partial_t( a \partial_t a)$.}
\begin{equation}
- \big[ \partial^2_t + 3 H \partial_t + 2 H^2 - \tfrac{\nabla^2}{a^2} \big]
C^{(0)}
=
8 \pi G M \, \tfrac{\delta^3({\bf x})}{a^3}
\;\; \Longrightarrow \;\;
C^{(0)}(t, {\bf x}) = - \frac{2 G M}{ a r}
\label{Ceqnsol0}
\end{equation}
\\
$\bullet \; 1^{st}order \; solution.$ It satisfies:
\begin{equation}
- \big[ \partial^2_t + 3 H \partial_t + 2 H^2 - \tfrac{\nabla^2}{a^2} \big]
C^{(1)}
\, \simeq \,
\kappa H^2 h^{(0)} C^{(0)}
\; , \label{Csol1}
\end{equation}
and equals:
\footnote{In (\ref{Ceqnsol1}) and thereafter, we use the fact 
that when $a(t)$ increases exponentially and a function $f(t)$ 
changes slowly, $\, \int^{t'}_0 dt'' a(t'') \, f(t'') \simeq
\frac{1}{H} a(t') \, f(t')$.}
\begin{equation}
C^{(1)}(t, {\bf x}) \, \simeq \, 
- \frac{1}{a} \int^t_0 \! dt' \frac{1}{a(t')}
\int^{t'}_0 \! dt'' a^2(t'') \, \kappa H^2 \, h^{(0)}(t'') \, C^{(0)}(t'')
\; , \label{Ceqnsol1}
\end{equation}
which implies the following (operator) equation for $C^{(1)}$:
\begin{equation}
Operator \; form \!: \quad
C^{(1)}(t, {\bf x}) \, \simeq \,
\frac{2 G M}{a r} \kappa H \int^t_0 \! dt' h^{(0)}(t')
\; . \label{Ceqnsol1operator}
\end{equation}
Because $\langle \Omega \vert h^{(0)} \vert \Omega \rangle = 0$, 
the vacuum expectation value of (\ref{Ceqnsol1operator}) is:
\begin{equation}
VEV \; form \!: \quad
\big\langle \Omega \big\vert \, C^{(1)}(t, {\bf x}) \,
\big\vert \Omega \big\rangle 
\, \simeq \, 0
\; . \label{Ceqnsol1vev}
\end{equation}
\\
$\bullet \; 2^{nd}order \; solution.$ Of the two possible ``source''
contributions that determine $C^{(2)}$, i.e. $h^{(1)} C^{(0)}$ and
$h^{(0)} C^{(1)}$, the first can only supply $\ln a$ terms. Hence,
we arrive at:
\begin{equation}
- \big[ \partial^2_t + 3 H \partial_t + 2 H^2 - \tfrac{\nabla^2}{a^2} \big]
C^{(2)}
\, \simeq \,
\kappa H^2 h^{(0)} C^{(1)}
\; . \label{Csol2}
\end{equation}
In view of (\ref{Ceqnsol1}):
\begin{eqnarray}
C^{(2)}(t, {\bf x}) 
\!\!& \simeq &
\nonumber \\
& \mbox{} & \hspace{-1.7cm}
- \frac{1}{a} \int^t_0 \! dt' \frac{1}{a(t')}
\int^{t'}_0 \! dt'' \, a^2(t'') \, \kappa^2 H^3 \, \tfrac{2 G M}{a(t') \, r} \,
h^{(0)}(t'') \int^{t''}_0 \! dt''' \, h^{(0)}(t''')
\; . \qquad \label{Ceqnsol2}
\end{eqnarray}

\newpage

We can now infer the (operator) equation for $C^{(2)}$:
\begin{eqnarray}
& \mbox{} & \hspace{-1.7cm}
Operator \, form \!:
\nonumber \\
& \mbox{} & \hspace{-1.7cm}
C^{(2)}(t, {\bf x}) \, \simeq \, 
- \frac{2 G M}{a(t) \, r} \, \kappa^2 H^2 \! \int^{t}_0 \! dt' \, h^{(0)}(t') \!
\int^{t'}_0 \! dt'' \, h^{(0)}(t'')
\; , \label{Ceqnsol2operator}
\end{eqnarray}
and subsequently the vacuum expectation value of (\ref{Ceqnsol2operator}):
\footnote{In (\ref{Ceqnsol2vevA}), we used that
$\langle \Omega \vert \, h^{(0)}(t') \, h^{(0)}(t'') \, \vert \Omega \rangle 
= -12 \tfrac{H^2}{4\pi^2} H t''$.
Various vacuum expectation values of graviton bilinears can be found 
in the Appendix.}
\begin{eqnarray}
\big\langle \Omega \big\vert \, C^{(2)}(t, {\bf x}) \,
\big\vert \Omega \big\rangle
\!\!& \simeq &\!\!
- \frac{2 G M}{a(t) \, r} \, \kappa^2 H^2 \int^{t}_0 \! dt' \int^{t'}_0 \! dt'' 
\big\langle \Omega \big\vert \, h^{(0)}(t') \, h^{(0)}(t'') \,
\big\vert \Omega \big\rangle 
\qquad \label{Ceqnsol2vevA} \\
\!\!& \simeq &\!\!
- \frac{2 G M}{a(t) \, r} \, \kappa^2 H^2 \times
\frac{1}{4\pi^2} \big[ -12 \, \tfrac16 (H t)^3 \big]
\; . \label{Ceqnsol2vevB}
\end{eqnarray}
More compactly, the vacuum expectation value of the first 
non-trivial correction to the Newtonian potential is:
\begin{eqnarray}
& \mbox{} & \hspace{-1.7cm}
VEV \; form \!: \quad
\nonumber \\
& \mbox{} & \hspace{-1.7cm} 
\big\langle \Omega \big\vert \, C^{(2)}(t, {\bf x}) \,
\big\vert \Omega \big\rangle
\, \simeq \,
C^{(0)} \Big[ \! -\frac{\kappa^2 H^2}{2\pi^2} \ln^3 a \Big] 
\, = \,
- \frac{2 G M}{a(t) \, r} \Big[ \! -\frac{\kappa^2 H^2}{2\pi^2} \ln^3 a \Big]
\; . \label{Ceqnsol2VEV} 
\end{eqnarray}

It is perhaps appropriate to make a few comments at this stage:
\\ [3pt]
- The result (\ref{Ceqnsol2VEV}) has the identical 
${\mathcal O}(\kappa^2)$ correction relative to the classical
Newtonian potential, with the similar correction obtained by
the explicit perturbative computation of \cite{Tan:2022xpn}
displayed in (\ref{tanresult}). Hence, the updated methodology 
and form of the LLOG equations passed a rather stringent test.
\\ [3pt]
- The Newtonian correction (\ref{Ceqnsol2VEV}) seemingly 
indicates a weakening of the gravitational force, a result
which perhaps can be understood from the term responsible 
for the dominant correction. Surprisingly, it is the 
``effective mass term'' $-2 {\widetilde H}^2 a^4
\sqrt{-{\widetilde g}}$, a term which increases with time. 
\\ [3pt]
- Although physical results require physical observables and
the current computations have been executed in a certain gauge,
while the result {\it may} persist in any gauge it most likely 
will {\it not} persist for physical observables.
\\ [3pt]
- One could argue that the underlying reason for achieving 
agreement between the LLOG equations and perturbation theory, 
is the presence in the quadratic 
operator of the above non-derivative term which effectively 
becomes a ``source'' term and ends up dominating. Apparently
the terms considered in \cite{Miao:2026cfl} were sub-dominant 
and could not agree with the perturbative calculation of the 
Newtonian correction in \cite{Tan:2022xpn}. 
\footnote{In particular, what we missed in \cite{Miao:2026cfl}
is that while $\frac{1}{\widetilde{D}_C} \, 1$ cannot lead to
secular growth, $\frac{1}{\widetilde{D}_C} \, \frac{1}{a}$ does.}
\\ [3pt]
- The triple logarithmic corrections that the Newtonian
potential receives at 1-loop order, need particular vertex
combinations in the perturbative computation of the 1-loop
self-energy. Each of the two vertices in the 1-loop diagram 
should be able to furnish an undifferentiated graviton 
field so that an undifferentiated graviton propagator can 
be formed and, thus, provide a logarithm. Such contributions 
are entirely stochastic and ultraviolet finite.
\\ [3pt]
- Nonetheless, it is still counter-intuitive that the dominant
term appears from expanding $\sqrt{-{\widetilde g}} \simeq
1 + \tfrac{\kappa}{2} h + \dots$ and, hence, involves a 
perturbative correction of the $C$ field, rather than a 
correction of the $A_{ij}$ dynamical graviton field; 
$h \equiv {\rm Tr}(h_{\mu\nu})$ is not dynamical.

\section{LLOG eqs: The 1-point function check}

\begin{figure}[H]
\centering
\includegraphics[width=3.9cm]{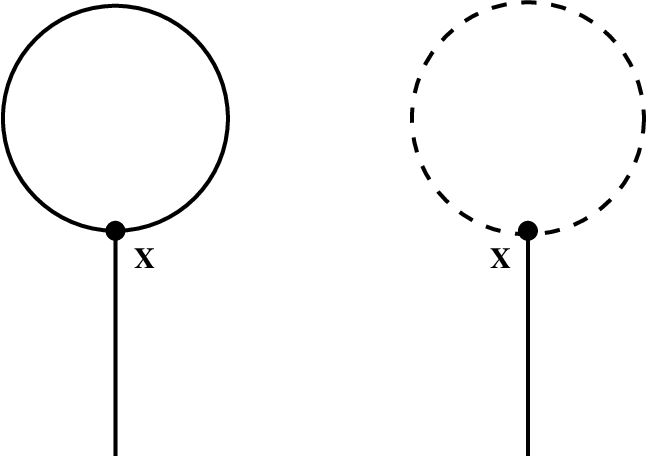}
\caption{\footnotesize The two diagrams contributing 
to the 1-loop 1PI 1-point function.
Graviton lines are solid whereas ghost lines are dashed.}
\label{1-pt_1-loop}
\end{figure}

The 1PI 1-point function is defined in the usual way:
\footnote{The functional variation of the action 
$i \frac{\delta S}{\delta h_{\mu\nu}(x)}$ at the vertex 
at $x$ gives the amputated 1-point function upon taking
the vacuum expectation value of the variation.}
\begin{equation}
\Big\langle \Omega \; \Big\vert \; 
i \, \frac{\delta S}{\delta h_{\mu\nu}(x)}
\; \Big\vert \; \Omega \Big\rangle
\Bigg\vert_{h^2, \, \overline{c} c}
\; . \label{1pt-def}
\end{equation}
The 1-loop diagrammatic representation is depicted 
in Figure \ref{1-pt_1-loop}.

With the understanding that $h_{\mu\nu}(x)$ is the external 
graviton of the 3-point vertex at $x$, we proceed to compute 
by firstly expanding the variation (\ref{1pt-def}) to 
${\mathcal O}(h^2)$ or ${\mathcal O}(\overline{c} c)$, and 
secondly by replacing every $h^2$ or $\overline{c} c$ with 
the corresponding propagator.

As an example, consider the graviton vertex $\#2$, one of 
the complete set of vertices shown in Table \ref{3point}
for the graviton and in Table \ref{ghost3point} for the
graviton-ghost:
\begin{equation}
\#2 \equiv 
i \kappa \, \mathcal{D} \, \big[ \tfrac14 h h^{\mu\nu} 
- \tfrac12 h^{\mu \rho} h^{\nu}_{~\rho} \big]
\; , \label{VertexNumber2}
\end{equation}
and perform the requisite reduction as well as go to $D=4$:
\begin{eqnarray}
\#2 &\!\! \longrightarrow \!\!& 
4 i \kappa \, \mathcal{D} \Big\{ \! -\tfrac12 \big[ 4 \, i \Delta_A (x ; x) 
+ i \Delta_B(x ; x) \big] {\overline \eta}^{\mu\nu}
\nonumber \\
& & \hspace{1.3cm}
+ \big[ \tfrac32 \, i \Delta_B (x ; x) + i \Delta_C (x ; x) \big]
\delta^{\mu}_{~0} \delta^{\nu}_{~0} \Big\}
\; , \label{reduction1} \\
& = &
\!\!\! - 4 i \kappa \, \frac{H^2}{8 \pi^2} \, \partial^{\alpha} 
a^2 \partial_{\alpha} \ln a \; {\overline \eta}^{\mu\nu}
= 
12 i \kappa \, \frac{H^4}{8 \pi^2} \, a^4 \, {\overline \eta}^{\mu\nu}
\; , \label{reduction2}
\end{eqnarray}
where we have used the relevant identities-formulae and 
Tables \ref{Limits}-\ref{Traces} from the Appendix.

\newpage

\begin{table}[h]
\setlength{\tabcolsep}{8pt}
\def\arraystretch{1.5}
\centering
\begin{tabular}{|@{\hskip 1mm }c@{\hskip 1mm }||c|}
\hline
$i$ & {\rm Graviton Vertex} \\
\hline\hline
1 & $\kappa \eta^{\mu\nu} \mathcal{D} [-\frac1{16} h^2 + \frac18
h^{\rho\sigma} h_{\rho\sigma} ]$ \\
\hline
2 & $\kappa \mathcal{D} [ \frac14 h h^{\mu\nu} - \frac12
h^{\mu \rho} h^{\nu}_{~\rho}$ ] \\
\hline
3 & $\kappa \partial_{\rho} \{ a^{D-2} h^{\rho\sigma} [-\frac12
h^{\mu\nu}_{~~ ,\sigma} + \frac14 \eta^{\mu\nu} h_{,\sigma}] \}$ \\
\hline
4 & $\kappa \eta^{\mu\nu} a^{D-2} [ \frac1{16} h^{,\rho} h_{,\rho}
- \frac18 h^{\rho\sigma , \alpha} h_{\rho \sigma , \alpha} ]$ \\
\hline
5 & $\kappa a^{D-2} [-\frac18 h^{,\mu} h^{,\nu} + \frac14
h^{\rho\sigma , \mu} h_{\rho\sigma}^{~~,\nu} ]$ \\
\hline
6 & $\kappa a^{D-2} [- \frac14 h^{\mu\nu , \rho} h_{,\rho} +
\frac12 h^{\mu \rho , \sigma} h^{\nu}_{~\rho , \sigma}]$ \\
\hline
7 & ${\scriptstyle (D-2)} \, \kappa H^2 a^D [ \frac14 \eta^{\mu\nu}
h_{0}^{~\rho} h_{0\rho} - \frac32 h_{0}^{~\mu} h_{0}^{~\nu} ]$ \\
\hline
8 & ${\scriptstyle (D-2)} \, \kappa H^2 a^D [\frac12 \delta^{(\mu}_{~~0}
h^{\nu)}_{~~0} h - 3 \delta^{(\mu}_{~~0} h^{\nu) \rho} h_{0\rho}]$ \\
\hline
9 & $\kappa \partial_{\rho} \partial_{\sigma} \{ a^{D-2} [- \frac14
\eta^{\mu\nu} h^{\rho \alpha} h^{\sigma}_{~\alpha} + \frac12
h^{\mu \rho} h^{\nu\sigma} ] \} + \kappa \partial_{\rho}
\partial^{(\mu} [a^{D-2} h^{\nu )}_{~~\sigma} h^{\rho\sigma} ]$ \\
\hline
10 & $\kappa \partial_{\rho} \{ a^{D-2} [\frac12 \eta^{\mu\nu}
h_{\alpha \sigma} h^{\rho \sigma , \alpha} -h_{\sigma}^{~(\mu}
h^{\nu) \rho ,\sigma} ] \} - 2 \kappa \partial^{(\mu} [a^{D-2}
h_{\rho\sigma} h^{\nu)\rho , \sigma}]$ \\
\hline
11 & ${\scriptstyle (D-2)} \, \kappa H a^{D-1} [-\frac12 h_{0}^{~(\mu}
h^{,\nu)} + h_{0}^{~(\mu} h^{\nu) \rho}_{~~~,\rho} + h_{0\rho}
h^{\rho (\mu ,\nu)} + h^{\rho (\mu} h_{0\rho}^{~~ ,\nu)} ]$ \\
\hline
12 & $\kappa a^{D-2} [\frac12 h^{\rho (\mu}_{~~~ ,\rho} h^{,\nu)} -
\frac12 h^{\rho (\mu ,\nu)} h_{,\rho} - h^{\rho (\mu}_{~~~ ,\rho}
h^{\nu) \sigma}_{~~~ ,\sigma} + h^{(\mu}_{~~ \rho ,\sigma}
h^{\nu) \sigma , \rho} ]$ \\
\hline
13 & ${\scriptstyle -(\frac{D}2 - 1)} \, \kappa H \eta^{\mu\nu}
\partial_{\rho} [a^{D-1} h^{\rho\sigma} h_{\sigma 0} ]$ \\
\hline
14 & ${\scriptstyle (\frac{D}2 - 1)} \, \kappa H a^{D-1} [ h_{0}^{~(\mu}
h^{,\nu)} + \delta^{(\mu}_{~~0} h^{\nu) \rho} h_{,\rho} ]$ \\
\hline
\end{tabular}
\caption{\footnotesize Distinguished 3-point graviton vertices
$\frac{\delta S}{\delta h_{\mu\nu}}$ in the new gauge in $D$
dimensions. Note that 
$\mathcal{D} \equiv \partial_{\rho} a^{D-2} \partial^{\rho}$.}
\label{3point}
\end{table}

The complete 1-loop perturbative computation is tedious but 
straightforward; the partial and total exact results are 
displayed in Table \ref{1pointresults}. Similarly, the LLOG 
results obtained by applying the updated {\it ``integrating
out''} rules of Section 3 are also displayed in Table
\ref{1pointresults}.

Finally, the agreement between the exact perturbative results
and the LLOG results provides a highly non-trivial check of
the updated rules described in Section 3. The full agreement
between the ``Exact results'' and the ``LLOG results'' of Table
\ref{1pointresults}, becomes apparent when we take into account
that to connect the ``LLOG results'' with the 1-point function
we must set $h_{\mu\nu} = 0 \, \Rightarrow \,
{\widetilde g}_{\mu\nu} = \eta_{\mu\nu}$.

\newpage

\begin{table}[H]
\setlength{\tabcolsep}{8pt}
\def\arraystretch{1.5}
\centering
\begin{tabular}{|@{\hskip 1mm }c@{\hskip 1mm }||c|}
\hline
$i$ & {\rm Graviton-Ghost Vertex} \\
\hline\hline
1 & $\kappa a^{D-2} \, {\overline c}^{\mu} \partial^2 c^{\nu}$ \\ 
\hline
2 & $-\kappa a^{D-2} \, {\overline c}^{\alpha} \partial^{\mu} 
\partial^{\nu} c_{\alpha}$ \\
\hline
3 & ${\scriptstyle (D-2)} \, \kappa H a^{D-1} \, {\overline c}^{\alpha} 
\delta^{(\mu}_{~~0} \, \partial^{\nu)} c_{\alpha}$ \\
\hline
4 & ${\scriptstyle -2(D-2)} \, \kappa H a^{D-1} \, {\overline c}^{(\mu} 
\partial^{\nu)} c^0$ \\
\hline
5 & ${\scriptstyle 2(D-2)} \, \kappa H^2 a^D \, {\overline c}^0 \, c^0
\delta^{\mu}_{~0} \, \delta^{\nu}_{~0}$ \\
\hline
6 & $-\kappa \, \partial^{(\mu} [ a^{D-2} \, {\overline c}^{\alpha}
c^{\nu}_{~,\alpha} ]$ \\
\hline
7 & $\tfrac12 \kappa \eta^{\mu\nu} \partial_{\alpha} 
[ \alpha^{D-2} \, {\overline c}^{\beta} c^{\alpha}_{~,\beta} ]$ \\
\hline
8 & $\kappa \partial^{\mu} \partial_{\alpha} [ a^{D-2} \,
{\overline c}^{\nu} \, c^{\alpha} ]$ \\
\hline
9 & $-\tfrac12 \kappa \eta^{\mu\nu} \partial_{\alpha} \partial_{\beta}
[ a^{D-2} \, {\overline c}^{\alpha} c^{\beta} ]$ \\
\hline
10 & ${\scriptstyle -(D-2)} \, \kappa H \delta^{(\mu}_{~~0} \, \partial_{\alpha}
[ a^{D-1} \, {\overline c}^{\nu)} c^{\alpha} ]$ \\
\hline
\end{tabular}
\caption{\footnotesize Distinguished 3-point graviton-ghost vertices 
$\frac{\delta S_{gh}}{\delta h_{\mu\nu}}$ in the new gauge and in 
$D$ dimensions.}
\label{ghost3point}
\end{table}

\begin{table}[H]
\setlength{\tabcolsep}{8pt}
\def\arraystretch{1.5}
\centering
\begin{tabular}{|@{\hskip 1mm }c@{\hskip 1mm }||c|c|}
\hline
{\rm Effective Lagrangian term} & {\rm Exact result} 
& {\rm LLOG result} \\
\hline\hline
$\mathcal{L}_{1+2+3}$ & $-\tfrac{13}{2} \eta^{\mu\nu}
+ 5 {\overline \eta}^{\mu\nu} + 3 \delta^{\mu}_{~0} \, \delta^{\nu}_{~0}$ 
& $-\tfrac{13}{2} {\widetilde g}^{\mu\nu} 
+ 5 {\overline \gamma}^{\mu\nu} + 3 u^{\mu} u^{\nu}$ \\ 
\hline
$\mathcal{L}_{4+5}$ & $-3 \delta^{\mu}_{~0} \, \delta^{\nu}_{~0}$ 
& $-3 u^{\mu} u^{\nu}$ \\
\hline
$\mathcal{L}_{6}$ & $+3 \eta^{\mu\nu}$ 
& $+3 N {\widetilde g}^{\mu\nu} + {\mathcal O}(h_{\alpha\beta})$ \\
\hline
$\mathcal{L}_{gh}$ & $-\tfrac32 \eta^{\mu\nu} 
- 4 \delta^{\mu}_{~0} \, \delta^{\nu}_{~0}$ 
& $-\tfrac32 {\widetilde g}^{\mu\nu} - 4 u^{\mu} u^{\nu}
+ {\mathcal O}(h_{\alpha\beta})$ \\
\hline\hline
$\mathcal{L}_{tot}$ & $+ \delta^{\mu}_{~0} \, \delta^{\nu}_{~0}$ 
& $+ u^{\mu} u^{\nu}$ \\

\hline
\end{tabular}
\caption{\footnotesize Results for the 1-loop 1PI 1-point function 
both from the exact perturbative computation and that from the LLOG
equations. All {\it exact} results should be multiplied by 
$\kappa \tfrac{H^4}{8\pi^2}a^4$ and all {\it LLOG} results should 
be multiplied by $\kappa \tfrac{{\widetilde H}^4}{8\pi^2}a^4
\sqrt{-{\widetilde g}}$.}
\label{1pointresults}
\end{table}

\section{Epilogue}

In one sentence, the requirement for LLOG computations to 
agree with exact perturbative results, in the case of the 
1-loop 1-point function steered us to realize that in addition
to differentiated graviton fields and ghost-antighost fields,
we must also ``integrate'' out undifferentiated constrained 
graviton components to derive the correct induced stress tensor.

Furthermore, without altering the functional form of the LLOG
equations the dominant contribution of the important constraint
degree of freedom $C$ was corrected via the necessity to concur
with the 1-loop force perturbative answer; schematically, the 
partial factor $\, \frac{1}{\widetilde{D}_C} \frac{1}{a} \,$ 
turned up to lead to the principal contribution.

The consistency checks on the LLOG equations 
(\ref{A_final}-\ref{C_final}) can and should be further 
enhanced by requiring full agreement with the exact 
perturbative result for the graviton 1-loop self-energy; 
an attainable elaborate computation.

Finally the LLOG equations should be solved in their own
validity region which, as expected, is much wider than
the perturbative one \cite{Miao:2025bmd}. De Sitter 
perturbation theory becomes inaccurate when 
$\langle \gamma^{ij} A_{ij} \rangle_{pert} \! \sim \! 
\kappa^2 H^2 \ln a$; beyond this time the {\it unknown} 
higher loop corrections are of the same order.
In contradistinction, the LLOG equations contain {\it all}
these higher corrections and can penetrate much further 
in time before becoming inaccurate:
$\langle \gamma^{ij} A_{ij} \rangle_{\scriptscriptstyle{LLOG}}
\! \sim \! \tfrac{1}{\kappa^2 H^2 \,}$.
Therefore, solving them would allow us to non-perturbatively 
re-compute physical quantities in their gauge fixing 
independent observable form.

\vspace{0.5cm}

\centerline{\bf Acknowledgements}

This work was partially supported by Taiwan NSTC grants
113-2112-M-006-013 and 114-2112-M-006-020, by NSF grant
PHY-2207514 and by the Institute for Fundamental Theory
at the University of Florida.


\newpage

\section{Appendix: Useful Identities \& Tables}

\subsection{Some relations from the 3+1 decomposition:}

\begin{eqnarray}
& \mbox{} &
{\widetilde H} = \frac{H}{N}
\quad , \quad 
{\widetilde H}^2 = \frac{H^2}{N^2} 
= - {\widetilde g}^{00} H^2 
\; , \label{Htilde} \\
& \mbox{} &
\delta^0_{\, \mu} = - \frac{1}{N} u_{\mu}
\quad , \quad
{\widetilde g}^{0\mu} = - \frac{1}{N} u^{\mu}
\quad , \quad
{\widetilde g}^{\mu\nu} u_{\mu} u_{\nu} 
= -1 =
{\widetilde g}_{\mu\nu} u^{\mu} u^{\nu} 
\; . \qquad\qquad \label{u's}
\end{eqnarray}

\subsection{Some tensor algebra identities:}

\begin{equation}
{\widetilde g}^{\mu\nu} =
\eta^{\mu\nu} - \kappa h^{\mu}_{\; \alpha} \; {\widetilde g}^{\alpha\nu}
\quad , \quad
{\widetilde g}^{\mu\nu} = {\overline \gamma}^{\mu\nu} - u^{\mu} u^{\nu}
\; , \label{ID1}
\end{equation}
\begin{equation}
\eta_{\mu\nu} =
{\overline \eta}_{\mu\nu} - \delta^0_{\; \mu} \delta^0_{\; \nu}
\quad , \quad
\delta^{\mu}_{\; \nu} =
{\overline \delta}^{\mu}_{\nu} + \delta^{\mu}_{\; 0} \delta^0_{\; \nu}
\; , \label{ID2}
\end{equation}
\begin{equation}
\delta^{\mu}_{\; \alpha} \delta^{\nu}_{\; \beta}
=
{\overline \delta}^{\mu}_{\; (\alpha} 
{\overline \delta}^{\nu}_{\; \beta)}
+ 2 \delta^{(\mu}_{\;\; 0} \, {\overline \delta}^{\nu)}_{\; (\alpha}
  \delta^0_{\; \beta)}
+ \delta^{\mu}_{\; 0} \delta^{\nu}_{\; 0}
  \delta^0_{\; \alpha} \delta^0_{\; \beta}
\; , \label{ID3}
\end{equation}
\begin{equation}
{\widetilde g}^{\mu\nu}_{~~ , \alpha} =
- {\widetilde g}^{\mu\rho} \, {\widetilde g}^{\nu\sigma}
{\widetilde g}_{\rho\sigma, \alpha} =
- {\widetilde g}^{\mu\rho} \, {\widetilde g}^{\nu\sigma}
\kappa h_{\rho\sigma , \alpha}
\; , \label{ID4}
\end{equation}
\begin{equation}
\partial_{\alpha} \sqrt{-{\widetilde g}} =
\tfrac12 \, \sqrt{-{\widetilde g}} \, {\widetilde g}^{\mu\nu}
{\widetilde g}_{\mu\nu , \alpha} = 
\tfrac12 \, \sqrt{-{\widetilde g}} \, {\widetilde g}^{\mu\nu}
\kappa h_{\mu\nu , \alpha}
\; , \label{ID5}
\end{equation}
\begin{equation}
\delta^{\mu}_{\; 0} =
{\widetilde g}^{\mu\rho} \, {\widetilde g}_{\rho 0} =
{\overline \gamma}^{\mu\rho} \, {\widetilde g}_{\rho 0}
- u^{\mu} u_0
\; , \label{ID6}
\end{equation}
\begin{equation}
\kappa h_{\mu 0} =
{\widetilde g}_{\mu 0} - \eta_{\mu 0} =
{\widetilde g}_{\mu 0} + \delta^0_{\; \mu} =
\tfrac12 u_0 B_{\mu} + u_0 u_{\mu} C 
\; , \label{ID7}
\end{equation}
\begin{equation}
\kappa h_{ij} = A_{ij} 
\quad , \quad
u^{\alpha} \kappa h_{\alpha\beta} =
- \tfrac12 B_{\beta} - u_{\beta} C
\quad , \quad
u^{\alpha} u^{\beta} \kappa h_{\alpha\beta} =
C
\; , \label{ID8}
\end{equation}
\begin{equation}
{\overline \gamma}^{\alpha\beta} \kappa h_{\alpha\beta} =
{\overline \gamma}^{\alpha\beta} A_{\alpha\beta}
\quad , \quad 
{\overline \gamma}^{\mu \alpha} A_{\alpha\beta} =
\Big[ {\widetilde g}^{\mu\alpha} + u^{\mu} u^{\alpha} \Big] 
A_{\alpha\beta} =
A^{\mu}_{\; \beta}
\; . \label{ID9}
\end{equation}
\begin{equation}
u \cdot \partial (A \!-\! \mathcal{A}) = 
\tfrac{\kappa^2 {\widetilde H}^3}{2 \pi^2} a \Big[ 
2 \!-\! \overline{\gamma}^{\alpha\beta} A_{\alpha\beta} \!-\! 4 C \Big] 
+ (\tfrac43 \!-\! 2 N^2) A^{\alpha\beta} \partial_{\alpha} B_{\beta} 
\; . \label{Atrace}
\end{equation}

\newpage

\subsection{Quadratic operator properties:} 

$\bullet \,$ 
The ${\widetilde D}_A$ quadratic operator equals:
\begin{equation}
{\widetilde D}_A 
\equiv
\partial_{\alpha} \big[ a^{D-2} {\sqrt {-\widetilde g}} \,
{\widetilde g}^{\alpha\beta} \partial_{\beta} \big]
\; , \label{D_A}
\end{equation}
and its operation on the three kinds of scalar propagators
gives:
\begin{eqnarray}
{\widetilde D}_A \, i {\widetilde \Delta}_A(x;x') 
&\!\!\!=\!\!\!& 
i \delta^D(x-x')
\; , \label{iDeltaA} \\
{\widetilde D}_A \, i {\widetilde \Delta}_B(x;x')
&\!\!\!=\!\!\!& 
i \delta^D(x-x') 
+ (D-2) {\widetilde H}^2 a^D {\sqrt {-\widetilde g}} \;
  i {\widetilde \Delta}_B(x;x')
\; , \label{iDeltaB} \\
{\widetilde D}_A \, i {\widetilde \Delta}_C(x;x')
&\!\!\!=\!\!\!& 
i \delta^D(x-x') 
+ 2(D-3) {\widetilde H}^2 a^D {\sqrt {-\widetilde g}} \; 
  i {\widetilde \Delta}_C(x;x') 
\; . \qquad \label{iDeltaC}
\end{eqnarray}
$\bullet \,$
The other two quadratic operators equal:
\begin{eqnarray}
{\widetilde D}_B 
&\!\!\!=\!\!\!&
{\widetilde D}_A - (D-2) \widetilde{H}^2 a^D \sqrt{-\widetilde{g}}
\; , \label{D_B} \\
{\widetilde D}_C 
&\!\!\!=\!\!\!& 
\widetilde{D}_A - 2(D-3) \widetilde{H}^2 a^D \sqrt{-\widetilde{g}}
\; . \label{D_C}
\end{eqnarray}
In $D \!=\! 4$ we have $(D-2) = 2(D-3) = 2$ so that 
${\widetilde D}_B = {\widetilde D}_C$ .
\\ [9pt]
\noindent
$\bullet \,$
The graviton quadratic operator equals: 
\begin{equation}
\widetilde{\mathcal D}^{\mu\nu\alpha\beta}
=
\tfrac12 \Big[ {\widetilde g}^{\mu(\alpha} \, {\widetilde g}^{\beta)\nu}
  - \tfrac12 {\widetilde g}^{\mu\nu} \, {\widetilde g}^{\alpha\beta} \Big] 
  \widetilde{\mathcal D}_A
+ (D-2){\widetilde H}^2 a^D \sqrt{- {\widetilde g}} \,
  u^{(\mu} \, {\widetilde g}^{\nu)(\alpha} \, u^{\beta)}
\; , \label{Dgrav}
\end{equation}
and allows us to check that indeed the proper condition
is satisfied:
\begin{equation}
\widetilde{\mathcal D}^{\mu\nu\alpha\beta} \, 
i \Bigl[ \mbox{}_{\alpha\beta} \widetilde{\Delta}_{\rho\sigma} \Bigr](x;x') 
=
\delta^{\mu}_{\; (\rho} \delta^{\nu}_{\; \sigma)} \,
i \delta^D (x-x')
\; . \label{propcheck}
\end{equation}

\noindent
$\bullet \,$
The ghost kinetic operator equals:
\begin{equation}
\widetilde{\mathcal D}^{\mu\alpha}
=
{\widetilde g}^{\mu\alpha} {\widetilde D}_A 
+ (D-2) {\widetilde H}^2 a^D {\sqrt {-\widetilde g}} \,
u^{\mu} u^{\alpha} 
\; , \label{ghostquadratic} 
\end{equation}
so that its action on (\ref{ghostprop}) gives:
\begin{equation}
\widetilde{\mathcal D}^{\mu\alpha} \, 
i \Bigl[ \mbox{}_{\alpha} \widetilde{\Delta}_{\rho} \Bigr](x;x') 
=
\delta^{\mu}_{\; \rho} \, i \delta^D (x-x')
\; . \label{ghostpropcheck}
\end{equation}

\subsection{Propagator coincidence limits in $D=4$ dimensions:}
\begin{table}[H]
\setlength{\tabcolsep}{8pt}
\def\arraystretch{1.5}
\centering
\begin{tabular}{|@{\hskip 1mm }c@{\hskip 1mm }||c|c|c|}
\hline
{\rm Limit} & $i\widetilde{\Delta}_{A}$ & $i\widetilde{\Delta}_{B}$ & 
$i\widetilde{\Delta}_{C}$ \\
\hline\hline
$i\widetilde{\Delta}_{I}(x;x')\vert_{x'=x}$ & 
$\tfrac{{\widetilde H}^2}{4\pi^2} \ln a + ``\infty"$ &
$-\tfrac{{\widetilde H}^2}{16\pi^2}$ & 
$+\tfrac{{\widetilde H}^2}{16\pi^2}$ \\
\hline
$\partial_{\mu} i\widetilde{\Delta}_{I}(x;x')\vert_{x'=x}$ & 
$-\tfrac{{\widetilde H}^3}{8\pi^2} \, a u_{\mu}$ & 
$0$ & 
$0$ \\
\hline
$\partial_{\mu} \partial_{\nu} i\widetilde{\Delta}_{I}(x;x')\vert_{x'=x}$ 
& $\tfrac{{\widetilde H}^4}{8\pi^2} \, a^2 
[ \frac34 \widetilde{g}_{\mu\nu} + u_{\mu} u_{\nu} ]$ 
& $-\tfrac{{\widetilde H}^4}{32\pi^2} \, a^2 \widetilde{g}_{\mu\nu}$ 
& $+\tfrac{{\widetilde H}^4}{32\pi^2} \, a^2 \widetilde{g}_{\mu\nu}$ \\
\hline
$\partial_{\mu} \partial'_{\nu} i\widetilde{\Delta}_{I}(x;x')\vert_{x'=x}$ 
& $-\tfrac{{3\widetilde H}^4}{32\pi^2} \, a^2 \widetilde{g}_{\mu\nu}$ 
& $+\tfrac{{\widetilde H}^4}{32\pi^2} \, a^2 \widetilde{g}_{\mu\nu}$ 
& $-\tfrac{{\widetilde H}^4}{32\pi^2} \, a^2 \widetilde{g}_{\mu\nu}$ \\
\hline
\end{tabular}
\caption{\footnotesize Coincidence limits of propagators 
and their derivatives in $D=4$ dimensions. In the above 
relations, we first take the derivatives and then the 
coincidence limit $x' \rightarrow x$.} 
\label{Limits}
\end{table}

\subsection{Tensor factor traces in $D=4$ dimensions:}
\begin{table}[H]
\setlength{\tabcolsep}{8pt}
\def\arraystretch{1.5}
\centering
\begin{tabular}{|@{\hskip 1mm }c@{\hskip 1mm }||c|c|c|c|}
\hline
{\rm Tensor\ Factor} & $[\mbox{}_{\mu\nu} \widetilde{T}^{\rho}_{~\rho}]$ 
& $[\mbox{}^{\rho}_{~\rho} \widetilde{T}^{\sigma}_{~\sigma}]$ 
& $[\mbox{}^{\rho}_{~\nu} \widetilde{T}_{\rho \sigma}]$ 
& $[\mbox{}^{\rho\sigma} \widetilde{T}_{\rho\sigma}]$ \\
\hline\hline
$[\mbox{}_{\mu\nu} \widetilde{T}^{A}_{~\rho\sigma}]$ 
& $-4 \overline{\gamma}_{\mu\nu}$ & $-12$ 
& $+2\overline{\gamma}_{\nu\sigma}$ & $+6$ \\
\hline
$[\mbox{}_{\mu\nu} \widetilde{T}^{B}_{~\rho\sigma}]$ 
& $0$ & $0$ & $+\overline{\gamma}_{\nu\sigma} - 3 u_{\nu} u_{\sigma}$ & $+6$ \\
\hline
$[\mbox{}_{\mu\nu} \widetilde{T}^{C}_{~\rho\sigma}]$ 
& $\!\!+2 \overline{\gamma}_{\mu\nu} \!+\! 2 u_{\mu} u_{\nu}\!\!$ & $+4$ 
& $\!\!+\overline{\gamma}_{\nu\sigma} \!-\! u_{\nu} u_{\sigma}\!\!$ & $+4$ \\
\hline
$\frac34 [\widetilde{T}^{A}] - \frac14 [\widetilde{T}^{B}] 
+ \frac14 [\widetilde{T}^{C}]$
& $\!\!-\frac52 \overline{\gamma}_{\mu\nu} \!+\! 
\frac12 u_{\mu} u_{\nu}\!\!$ & $-8$ 
& $\!\!+\frac32 \overline{\gamma}_{\nu\sigma} \!+\! 
\frac12 u_{\nu} u_{\sigma}\!\!$ & $+4$ \\
\hline
\end{tabular}
\caption{\footnotesize Traces of tensor factors in $D=4$ dimensions.}
\label{Traces}
\end{table}

\newpage


\end{document}